\documentclass[default,iicol]{class/sn-jnl}% Default with double column layout

\usepackage{graphicx}%
\usepackage{multirow}%
\usepackage{amsmath,amssymb,amsfonts}%
\usepackage{amsthm}%
\usepackage{mathrsfs}%
\usepackage[title]{appendix}%
\usepackage{xcolor}%
\usepackage{textcomp}%
\usepackage{manyfoot}%
\usepackage{booktabs}%
\usepackage{algorithm}%
\usepackage{algorithmicx}%
\usepackage{algpseudocode}%
\usepackage{listings}%
\usepackage{xspace}
\usepackage{tabularx}
\usepackage{makecell}
\usepackage{caption}
\usepackage{subcaption}
\usepackage{placeins}
\usepackage{bookmark}
\usepackage[shortlabels]{enumitem}
\usepackage{orcidlink}

\newcommand\third{0.325}
\newcommand\half{0.495}

\newcommand\halfwidth{0.475}

\def \fwhm {\ensuremath{\text{FWHM}}\xspace}
\def \fwhmgen {\ensuremath{\text{FWHM}_{\text{gen}}}\xspace}
\def \fwhmdep {\ensuremath{\text{FWHM}_{\text{dep}}}\xspace}
\def \res {\ensuremath{\text{FWHM}_{\text{gen}}/2.355}\xspace}
\def \gravnet {GravNet\xspace}
\def \belletwo {Belle\,II\xspace}
\def \superkekb {SuperKEKB\xspace}
\def \thetagen {\ensuremath{\theta_{\mathrm{gen}}}\xspace}
\def \phigen {\ensuremath{\phi_{\mathrm{gen}}}\xspace}
\def \edep {\ensuremath{E_{\mathrm{dep}}}\xspace}
\def \egen {\ensuremath{E_{\mathrm{gen}}}\xspace}
\def \erecgnn {\ensuremath{E_{\mathrm{rec}}^{\mathrm{GNN}}}\xspace}
\def \erecbasf {\ensuremath{E_{\mathrm{rec}}^{\mathrm{basf2}}}\xspace}
\def \erecrawbasf {\ensuremath{E_{\mathrm{rec,\, raw}}^{\mathrm{basf2}}}\xspace}

\def \ereccrystal {\ensuremath{E^\mathrm{crystal}_\mathrm{rec}}\xspace}
\def \treccrystal {\ensuremath{t^\mathrm{crystal}_\mathrm{rec}}\xspace}

\def \pyg {\texttt{PyTorch Geometric}\xspace}

\def \basf {\texttt{basf2}\xspace}
\def \etagen {\ensuremath{\eta_\text{gen}}\xspace}
\def \etadep {\ensuremath{\eta_\text{dep}}\xspace}

\newcommand{\gev}{\ensuremath{\mathrm{\,Ge\kern -0.1em V}}\xspace}
\newcommand{\mev}{\ensuremath{\mathrm{\,Me\kern -0.1em V}}\xspace}

\usepackage[normalem]{ulem}

\usepackage{etoolbox}   
\makeatletter
\patchcmd{\@maketitle}{\artauthors}{{\artauthors}}{}{}
\makeatother

\begin{document}

\title[Article Title]{Photon Reconstruction in the \belletwo Calorimeter\newline Using Graph Neural Networks}

\author{F.~Wemmer\,\orcidlink{0000-0002-6475-0834}}

\author{I.~Haide\,\orcidlink{0000-0003-0962-6344}}

\author{J.~Eppelt\,\orcidlink{0000-0001-8368-3721}}

\author{T.~Ferber\,\orcidlink{0000-0002-6849-0427}}

  \author{A.~Beaubien\,\orcidlink{0000-0001-9438-089X}} % 6683

  \author{P.~Branchini\,\orcidlink{0000-0002-2270-9673}} % 2577

  \author{M.~Campajola\,\orcidlink{0000-0003-2518-7134}} % 5223

  \author{C.~Cecchi\,\orcidlink{0000-0002-2192-8233}} % 2433

  \author{P.~Cheema\,\orcidlink{0000-0001-8472-5727}} % 15264

  \author{G.~De~Nardo\,\orcidlink{0000-0002-2047-9675}} % 2459

  \author{C.~Hearty\,\orcidlink{0000-0001-6568-0252}} % 2450

  \author{A.~Kuzmin\,\orcidlink{0000-0002-7011-5044}} % 2520

  \author{S.~Longo\,\orcidlink{0000-0002-8124-8969}} % 2396

  \author{E.~Manoni\,\orcidlink{0000-0002-9826-7947}} % 2305

  \author{F.~Meier\,\orcidlink{0000-0002-6088-0412}} % 3103
  
  \author{M.~Merola\,\orcidlink{0000-0002-7082-8108}} % 2456

  \author{K.~Miyabayashi\,\orcidlink{0000-0003-4352-734X}} % 2327

  \author{S.~Moneta\,\orcidlink{0000-0003-2184-7510}} % 13303

  \author{M.~Remnev\,\orcidlink{0000-0001-6975-1724}} % 2785

  \author{J.~M.~Roney\,\orcidlink{0000-0001-7802-4617}} % 2244

  \author{J.-G.~Shiu\,\orcidlink{0000-0002-8478-5639}} % 2412

  \author{B.~Shwartz\,\orcidlink{0000-0002-1456-1496}} % 2122

  \author{Y.~Unno\,\orcidlink{0000-0003-3355-765X}} % 2420

  \author{R.~van~Tonder\,\orcidlink{0000-0002-7448-4816}} % 2706

  \author{R.~Volpe\,\orcidlink{0000-0003-1782-2978}} % 20183

\abstract{We present the study of a fuzzy clustering algorithm for the \belletwo electromagnetic calorimeter using Graph Neural Networks.
We use a realistic detector simulation including simulated beam backgrounds and focus on the reconstruction of both isolated and overlapping photons.
We find significant improvements of the energy resolution compared to the currently used reconstruction algorithm for both isolated and overlapping photons of more than 30\% for photons with energies $E_{\gamma}<0.5\,\gev$ and high levels of beam backgrounds.
Overall, the GNN reconstruction improves the resolution and reduces the tails of the reconstructed energy distribution and therefore is a promising option for the upcoming high luminosity running of \belletwo.}


\maketitle

% one section per include
\section{Introduction}\label{sec:intro}
The \belletwo experiment is located at the high-intensity, asymmetric electron-positron-collider \superkekb in Tsukuba, Japan.
\superkekb is colliding 4\,\gev  positron  and  7\,\gev  electron beams at a center-of-mass energy of around 10.58\,\gev to search for rare meson decays and new physics phenomena.
Many of these decays include photons in the final state that are reconstructed exclusively in the electromagnetic calorimeter.
The experimental program of \belletwo targets a significantly increased instantaneous luminosity that ultimately exceeds the predecessor experiment by a factor of 30.
This increase in luminosity also leads to a significant increase in beam-induced backgrounds~\cite{Natochii:2022vcs}.
These background processes produce both high-energy particle interactions that could be misidentified as physics signals, but also energy depositions of low-energy particles that degrade the energy resolution of the electromagnetic crystal calorimeter.
The electronics signals from the calorimeter are interpreted during a process called reconstruction to determine the properties of particles that created the signals.

In this paper, we describe a fuzzy clustering algorithm based on Graph Neural Networks~(GNNs) to reconstruct photons.
The term fuzzy clustering~\cite{fuzzy} refers to the partial assignment of individual calorimeter crystals to several clustering classes.
In our case, these are potentially overlapping, different signal photons, but also a beam background class.

The paper is organized as follows: Section\,\ref{sec:related} gives an overview of related work on Machine Learning for calorimeter reconstruction.
Section\,\ref{sec:ecl} describes the \belletwo electromagnetic calorimeter. 
The event simulation and details of the beam background simulation are discussed in Section\,\ref{sec:dataset}.
The conventional \belletwo reconstruction algorithm and the new GNN algorithm are described in Section\,\ref{sec:reconstruction}.
We introduce the metrics used to measure the performance of the GNN algorithm in Section\,\ref{sec:metrics}.
The main performance studies and results are discussed in Section\,\ref{sec:results}.
We summarize our results in Section\,\ref{sec:conclusion}.

\section{Related work}\label{sec:related}
Machine Learning is widely used in high energy physics for the reconstruction of calorimeter signals both for clustering\,\cite{canudas2022graph, Valsecchi_2023}, energy regression\,\cite{ereg, Belayneh2019CalorimetryWD}, but also particle identification\,\cite{Boldyrev_2020, Charan:2023ldg} and fast simulation\,\cite{Paganini:2017dwg, Buhmann:2020pmy, ATLAS:2022jhk}.
Most of the recent work has been performed in the context of the high-granularity calorimeter (HGCAL) at CMS~\cite{hgcal1, hgcal2}.
For \belletwo, the use of machine learning utilizing the electromagnetic calorimeter is so far limited to image-based particle identification in the barrel~\cite{Novosel:2023cki, Charan:2023ldg}.

GNNs are now widely recognized as one possible solution for irregular geometries in high energy physics\,\cite{Shlomi:2020gdn, Duarte:2020ngm, DeZoort:2023vrm}.
GNN architectures that are able to learn a latent space representation of the detector geometry itself~\cite{Wang:2018nkf, Qasim:2019otl} are the basis of the work presented in this paper.

Previous work has focused on simplified and idealized detector geometries, often approximated as a regular grid of readout cells expressed as 2D or 3D images.
Additionally, the presence of geometry changes and overlaps between barrel and endcap regions, large variations of cell sizes, and the presence of very high spatially non-uniform noise levels induced by beam background energy depositions are neglected.

For a complete list of works in particle physics that utilize machine learning, we refer to the review \cite{hepmllivingreview}.

\section{The Belle~II Electromagnetic Calorimeter}\label{sec:ecl}

The \belletwo detector consists of several subdetectors arranged around the beam pipe in a cylindrical structure that is described in detail in Ref.~\cite{Kou:2018nap, Belle-II:2010dht}.
We define the $z$-axis of the laboratory frame as the central axis of the solenoid.
The positive direction is pointing in the direction of the electron beam.
The $x$ axis is horizontal and points away from the accelerator center, while the $y$ axis is vertical and points upwards.
The longitudinal direction, the transverse plane with azimuthal angle $\phi$, and the polar angle $\theta$ are defined with respect to the detector’s solenoidal axis.

The \belletwo electromagnetic calorimeter~(ECL) consists of 8736 Thallium-doped CsI (CsI(Tl)) crystals that are grouped in a forward endcap, covering a polar angle $12.4^{\circ} < \theta < 31.4^{\circ}$, a barrel, covering a polar angle $32.2^{\circ} < \theta < 128.7^{\circ}$, and a backward endcap, covering a polar angle $130.7^{\circ} < \theta < 155.1^{\circ}$.
The crystals have a trapezoidal geometry with a nominal cross-sectional area of approximately $6 \times 6$ cm$^2$ and a length of 30~cm, providing 16.1 radiation lengths of material.
While crystals in the barrel are similar in cross-section and shape, the crystals in the endcaps vary with masses between 4.03\,kg and 5.94\,kg~\cite{ikeda:1999}; crystals in the endcaps also have significantly more passive material in front of the crystals.
Each crystal is aligned in the direction of the collision point with a small tilt in polar angle $\theta$ to reduce detection inefficiencies from particles passing between two crystals.
Crystals in the barrel additionally have a small tilt in azimuthal angle $\phi$.  
The scintillation light produced in the CsI(Tl) crystals is read out by two photodiodes glued to the back of each crystal.
After shaping electronics, the waveform is digitized and the crystal energy \ereccrystal over baseline and time \treccrystal since trigger time of the energy deposition are reconstructed online using FPGAs~\cite{Aulchenko_2017}.
Waveforms of crystals with energy depositions above 50\,MeV are stored for offline processing to allow for electromagnetic vs. hadronic shower identification through pulse shape discrimination (PSD)~\cite{longo:2020164562}.
Available information from PSD is
\begin{itemize}
    \item the fit type ID of a multi-template fit indicating which of the possible templates provides the best goodness-of-fit,
    \item the respective $\chi^2$ value as an indicator of the goodness-of-fit,
    \item and the ratio of reconstructed hadronic and photon template energies, referred to as PSD hadronic energy ratio in the following.
\end{itemize}

\section{Data Set}\label{sec:dataset}

In this work, we use simulated events to train and evaluate the reconstruction algorithms. 
The detector geometry and interactions of final-state particles with detector materials are simulated using \textsc{Geant4}\,\cite{Agostinelli:2002hh} combined with a dedicated detector response simulation. 
Simulated events are reconstructed and analyzed using the \belletwo Analysis Software Framework~(\basf)\,\cite{basf21, basf22}.
We simulate isolated photons, with energy \hbox{$0.1 < \egen < 1.5\,\text{GeV}$}, and direction \hbox{$17^{\circ} < \thetagen < 150^{\circ}$} and \hbox{$0^{\circ} < \phigen < 360^{\circ}$} drawn randomly from independent uniform distributions in $E$, $\theta$, and $\phi$. 
The generation vertex of the photons is $x=0$, $y=0$, and $z=0$.
For events with two overlapping photons, we first draw randomly one photon with independent uniform distributions as outline above.
We then simulate a second photon with an angular separation $2.9 < \Delta\alpha < 9.7\,^{\circ}$ drawn randomly from uniform distributions in $\Delta\alpha$ and in $E$.
This angular separation covers approximately the distance needed to create two overlapping clusters.
These two cases are typical calorimeter signatures in \belletwo that describe the majority of photons.
We note that the reconstructions of hadrons is a more difficult task not yet covered by our algorithm.

As part of the simulation, we overlay simulated beam background events corresponding to different collision conditions to our signal particles~\cite{Liptak:2021tog, Natochii:2022vcs}.
The simulated beam backgrounds correspond to an instantaneous luminosity of \hbox{$\mathcal{L}_{\text{beam}}=1.06\times10^{34}$\,cm$^{-2}$s$^{-1}$} (called \textit{low beam background}), and \hbox{$\mathcal{L}_{\text{beam}}=8\times10^{35}$\,cm$^{-2}$s$^{-1}$} (called \textit{high beam background}).
Those two values approximately correspond to the conditions in 2021, and the expected conditions slightly above the design luminosity, respectively.
The spatial distribution of beam backgrounds is asymmetric: They are much higher in the backward endcap than in the forward endcap, and they are slightly higher in the barrel than in the forward endcap.
Additional electronics noise per crystal of about 0.35\,MeV is included in our simulation as well.

The supervised training and the performance evaluation both use labeled information that relies on matching reconstructed information with the simulated \textit{truth} information. 
For each of the four configurations, isolated and overlapping photons with low and high beam backgrounds, we use 1.8~million events for training and 200\,000 events for validation. 
The performance evaluation is carried out on a large number of statistically independent samples simulated with various energies and in different detector regions.

We then study the performance of the GNN clustering algorithm in all four scenarios and compare it to the baseline \basf reconstruction. Both reconstruction algorithms are described in detail in Sec.~\ref{sec:reconstruction}.

\subsection{Isolated Photon}\label{para:one_cluster}
\begin{figure*}[ht]
     \centering
     \begin{subfigure}[t]{\third\textwidth}
         \centering
         \includegraphics[width=\textwidth]{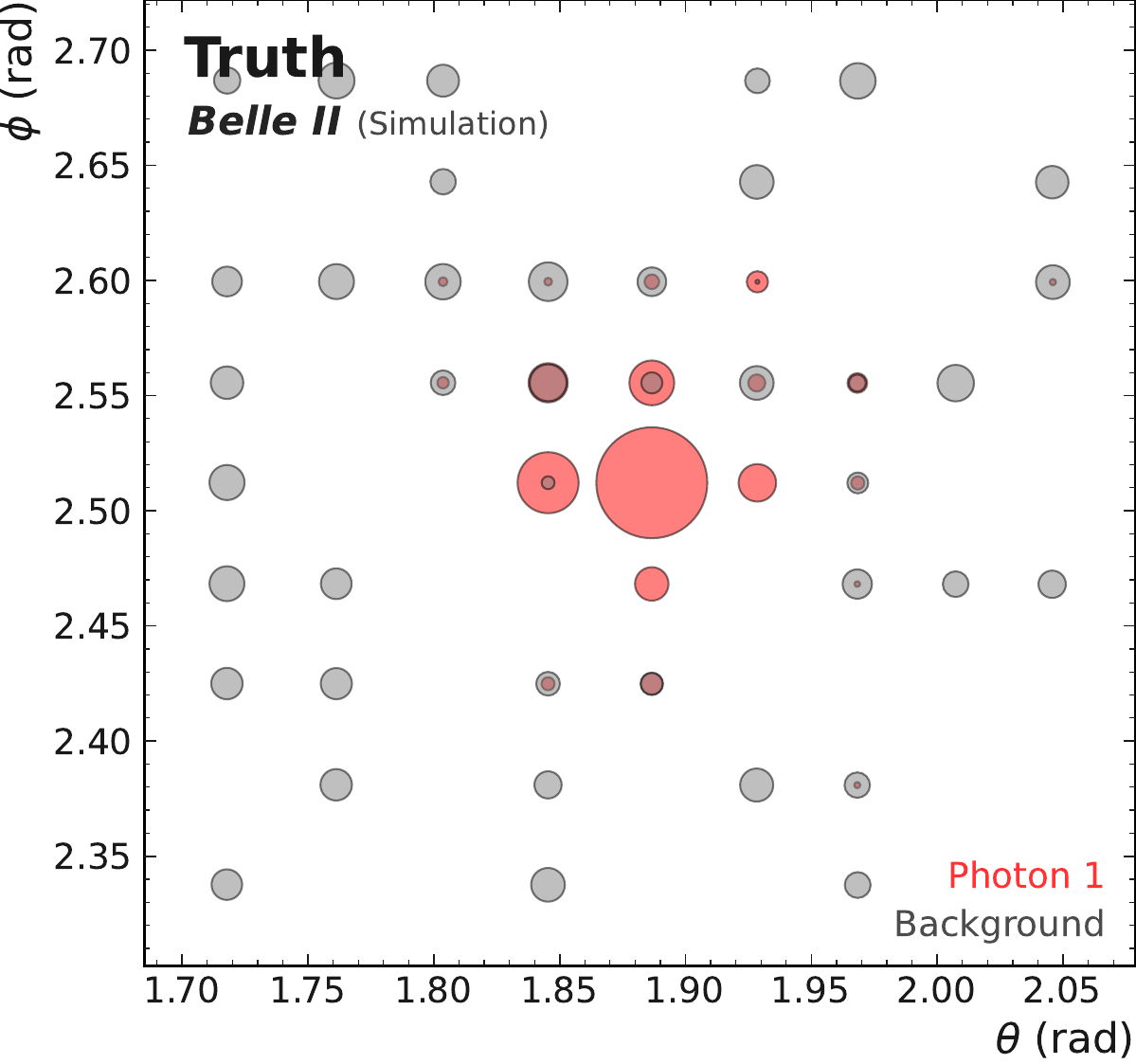}
         % \caption{Truth}
         \label{fig:one_truth}
     \end{subfigure}
     \hfill
     \begin{subfigure}[t]{\third\textwidth}
         \centering
         \includegraphics[width=\textwidth]{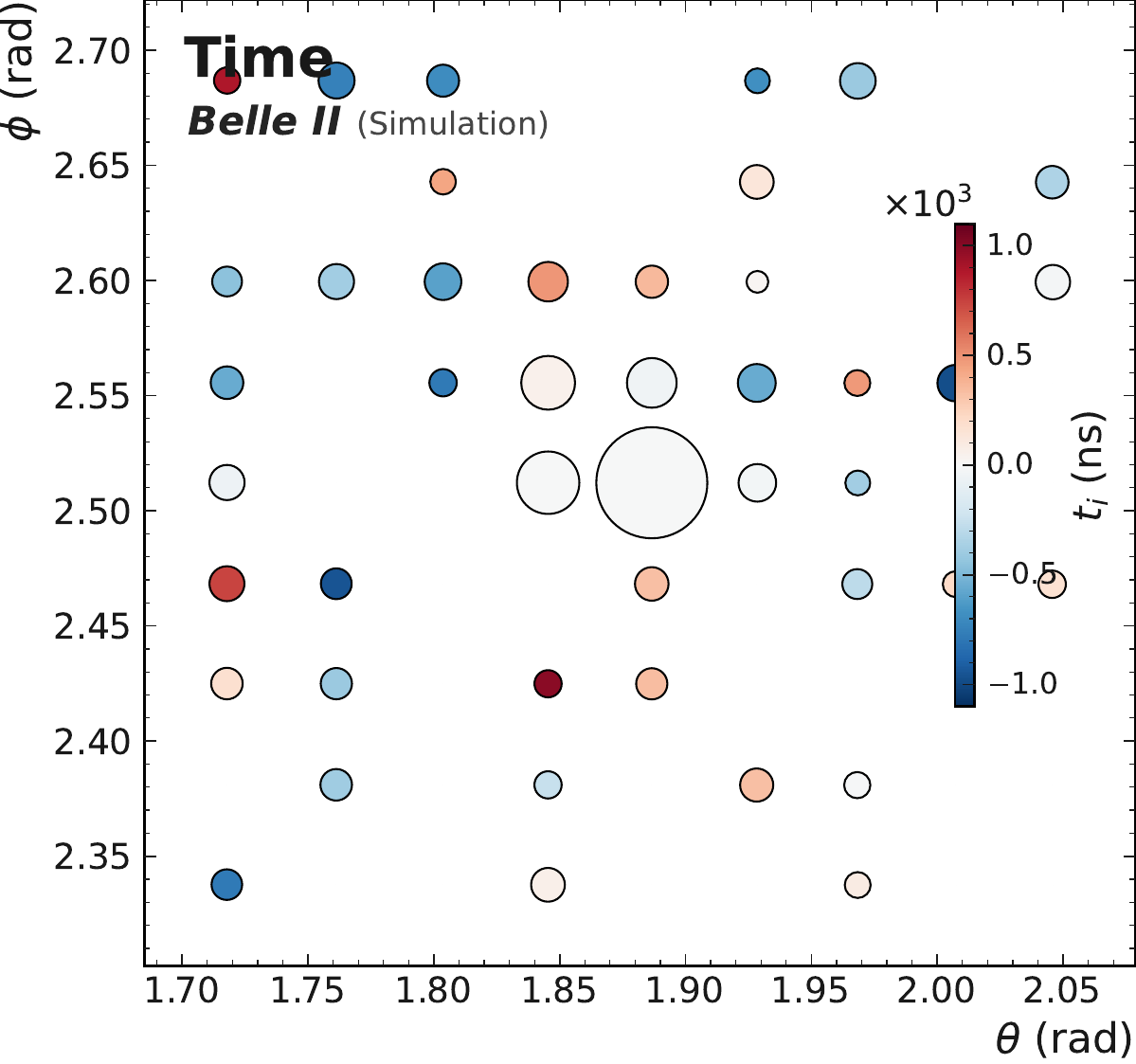}
         % \caption{Time}
         \label{fig:one_time}
     \end{subfigure}
     \hfill
     \begin{subfigure}[t]{\third\textwidth}
         \centering
         \includegraphics[width=\textwidth]{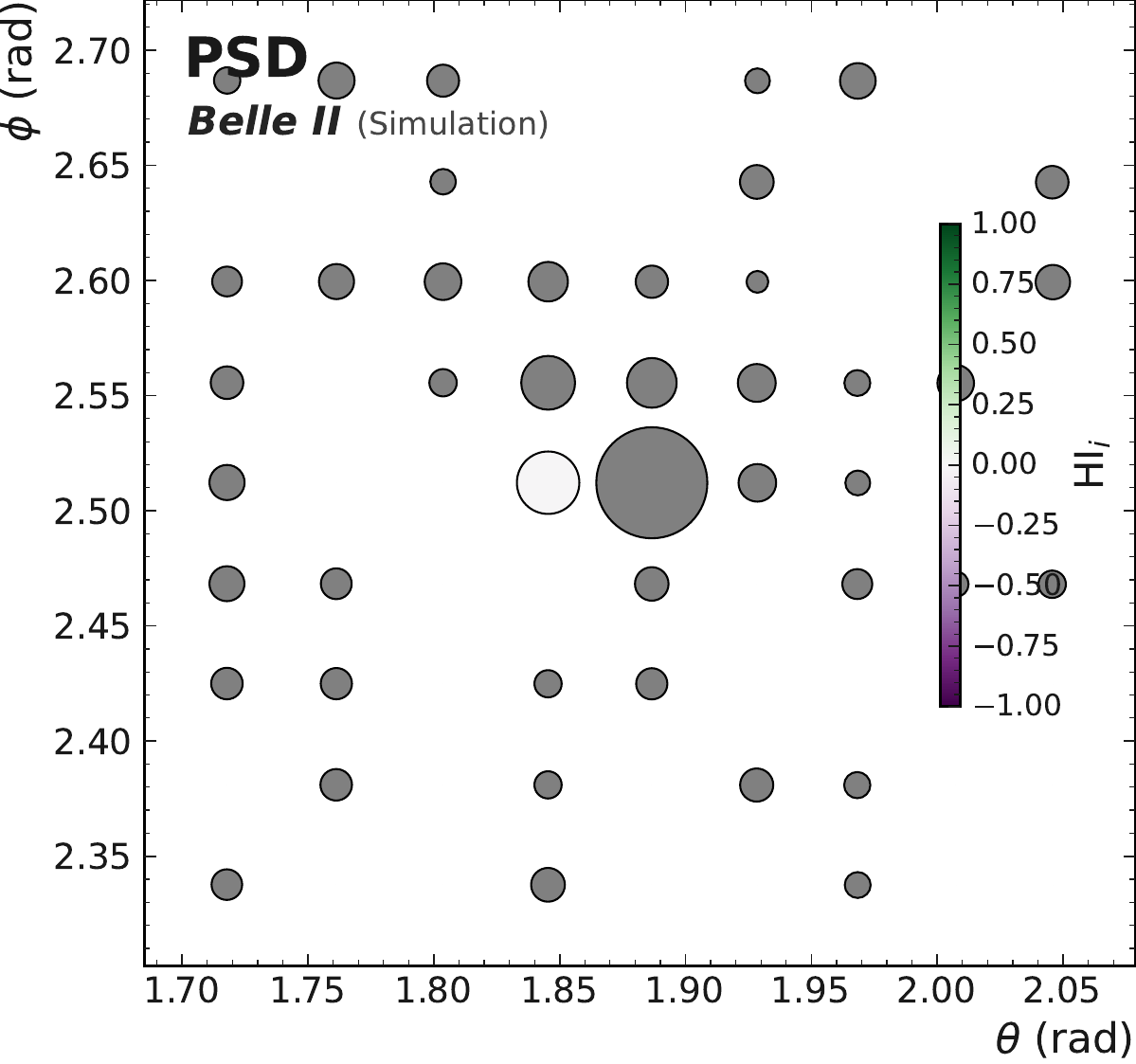}
         % \caption{PSD}
         \label{fig:one_psd}
     \end{subfigure}\\
     
     \begin{subfigure}[t]{\third\textwidth}
         \centering
         \includegraphics[width=\textwidth]{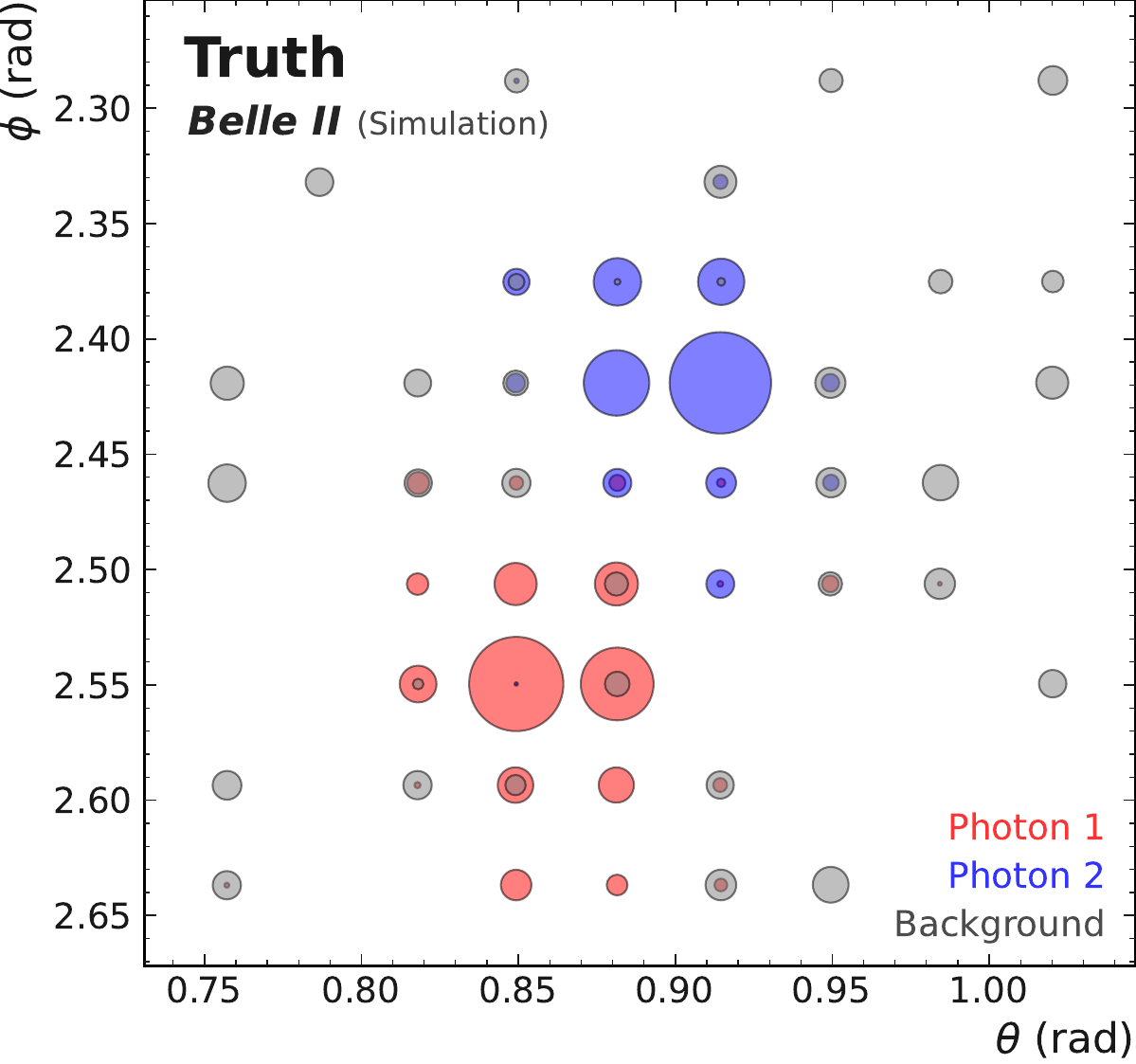}
         \caption{Truth assignment, colors indicate the fraction belonging to each of the photons and beam background.}
         \label{fig:two_truth}
     \end{subfigure}
     \hfill
     \begin{subfigure}[t]{\third\textwidth}
         \centering
         \includegraphics[width=\textwidth]{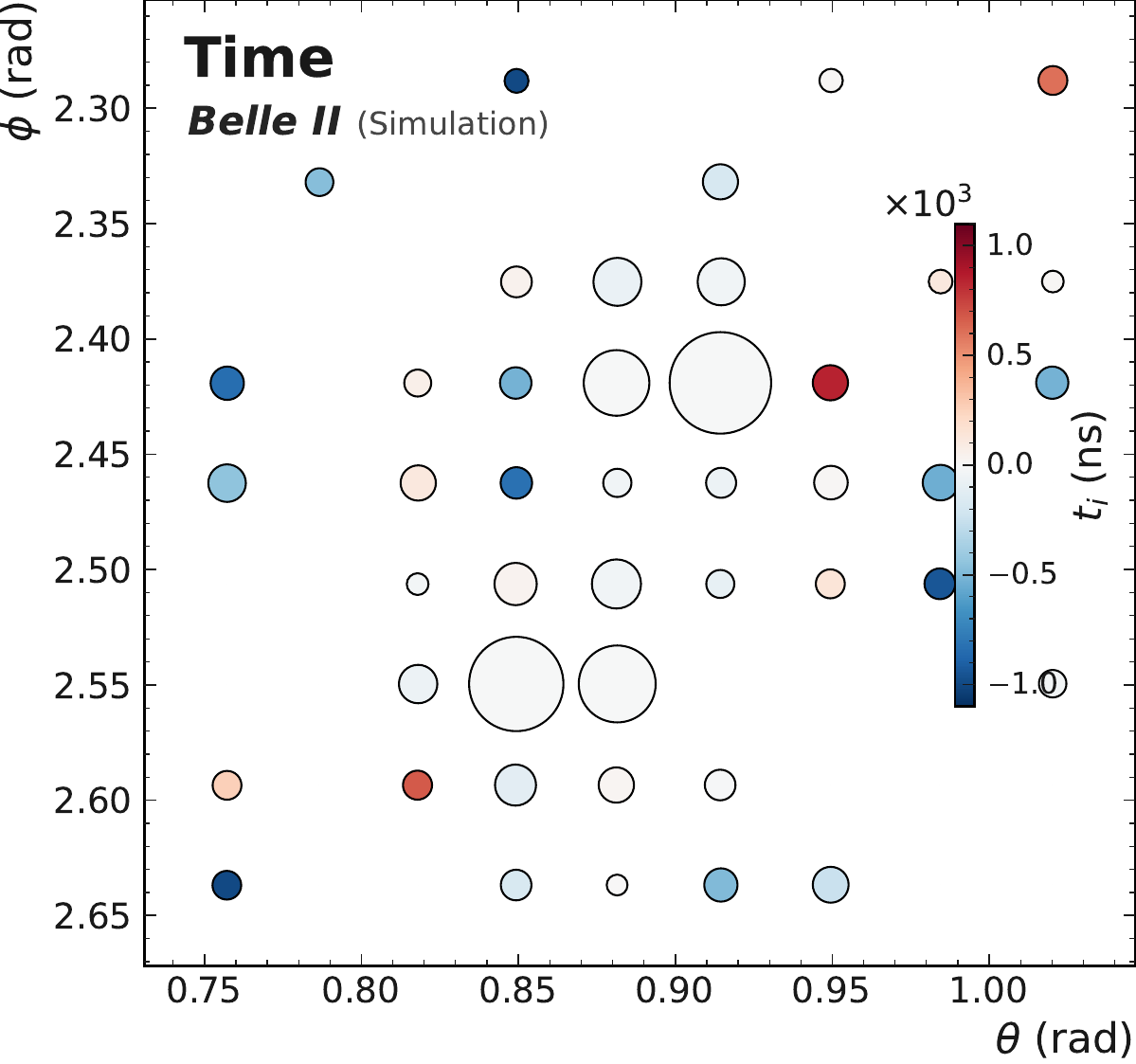}
         \caption{Reconstructed time $t$ since trigger time.}
         \label{fig:two_time}
     \end{subfigure}
     \hfill
     \begin{subfigure}[t]{\third\textwidth}
         \centering
         \includegraphics[width=\textwidth]{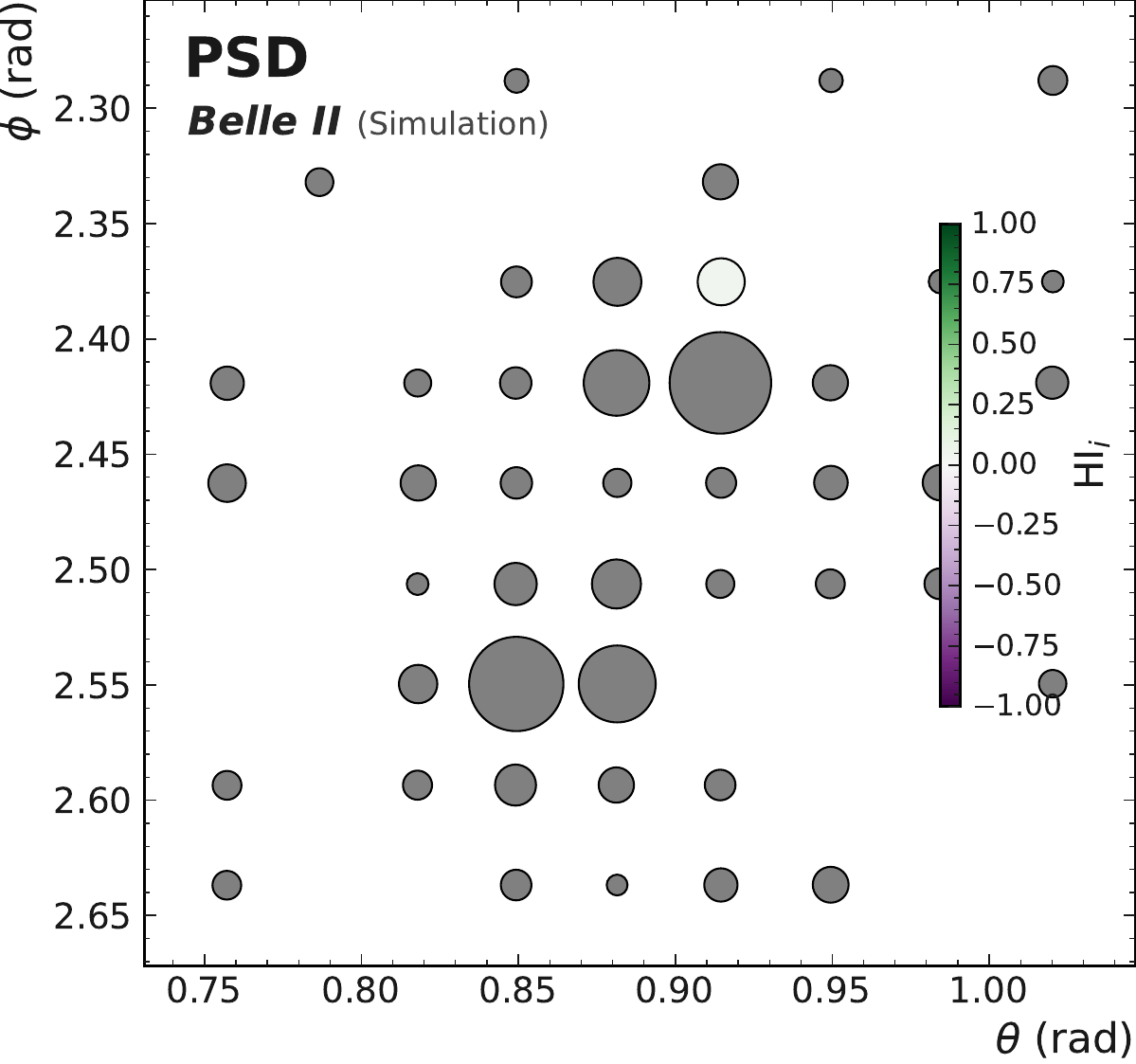}
         \caption{Reconstructed PSD hadronic energy ratio. Gray markers indicate that no PSD information is available.}
         \label{fig:two_psd}
     \end{subfigure}
    \caption{Typical event displays showing (left) simulated truth assignments, (center) input variables time, and (right) PSD hadronic energy ratio for (top) isolated and (bottom) overlapping photons for two example events with high beam background. The marker centers indicate the crystal centers, the marker area is proportional to the truth energy deposition for the left plots; it is proportional to the reconstructed crystal energy for the other plots.}
    \label{fig:edtwonominal}
\end{figure*}

To study isolated photons, we use the simulated events with a generated isolated photon only.
For each event, we select a region of interest (ROI): 
We first determine the azimuthal angles of the fourth neighbour on either side of the local maximum (LM), and the polar angles of the fourth neighbours on either direction of the LM.
We then include all crystals in that angular range.
In the barrel this defines a regular $9\times9$ array of crystals centered around a LM, while in the endcaps this array is not necessarily regular, but can contain a few crystals more or less.
The LM is a crystal with at least 10\,MeV of reconstructed crystal energy, and energy higher than all its direct eight neighbors. 
The LM must be the only LM in the ROI, and the matched truth particle must be a simulated photon responsible for at least 20\% of the reconstructed crystal energy. 
Precisely, for the LM we require the ratio
\begin{equation}r^{\gamma_1}_\mathrm{LM}=\frac{E^{\gamma_1\mathrm{,crystal}_\mathrm{LM}}_\mathrm{dep}}{E^{\mathrm{crystal}_\mathrm{LM}}_\mathrm{rec}}\geq0.2.
\end{equation}
Here, $E^{\gamma_1\mathrm{,crystal}_\mathrm{LM}}_\mathrm{dep}$ denotes the truth energy deposition of photon~1 in the LM, and $E^{\mathrm{crystal}_\mathrm{LM}}_\mathrm{rec}$ the reconstructed crystal energy in the LM.
The crystals contained in the ROI are considered for the clustering by the GNN algorithm and significantly extend the $5\times5$ area considered by the baseline algorithm (Sec.~\ref{sec:reconstruction}). 
Furthermore, the ROI represents the area of the local coordinate system later used as an input feature, with the LM as the origin.
Figure~\ref{fig:edtwonominal}~(top) shows a typical isolated photon event with high beam background.

\subsection{Overlapping Photons}\label{para:two_cluster}
Two different photons that deposit some of their energy in identical crystals are referred to as overlapping photons.
To study overlapping photons, we use the simulated events with two overlapping photons only.
We select events that have exactly two LMs that must fulfill the following selection criteria:
\begin{enumerate}[a),itemsep=0.3em, after=\vspace{0.3em}]    % modified spacing to make list less crowded
    \item each LM must have reconstructed crystal energies greater than 10~MeV,
    \item $r^{\gamma_1}_\mathrm{LM_1}\geq0.2$ and $r^{\gamma_1}_\mathrm{LM_1}>r^{\gamma_2}_\mathrm{LM_1}$,
    \item $r^{\gamma_2}_\mathrm{LM_2}\geq0.2$ and $r^{\gamma_2}_\mathrm{LM_2}>r^{\gamma_1}_\mathrm{LM_2}$. 
\end{enumerate}
We refer to criteria a)-c) as LM \textit{separation criteria} since they ensure that the particles form two separate LMs. Additionally, events must meet the \textit{overlap criterion}: 
\begin{enumerate}[d)]
    \item each of the two photons must deposit at least 10~MeV energy in shared crystals within a $5\times5$ area around its respective LM.
\end{enumerate}
\begin{figure}
\centering 
    \includegraphics[width=\halfwidth\textwidth]{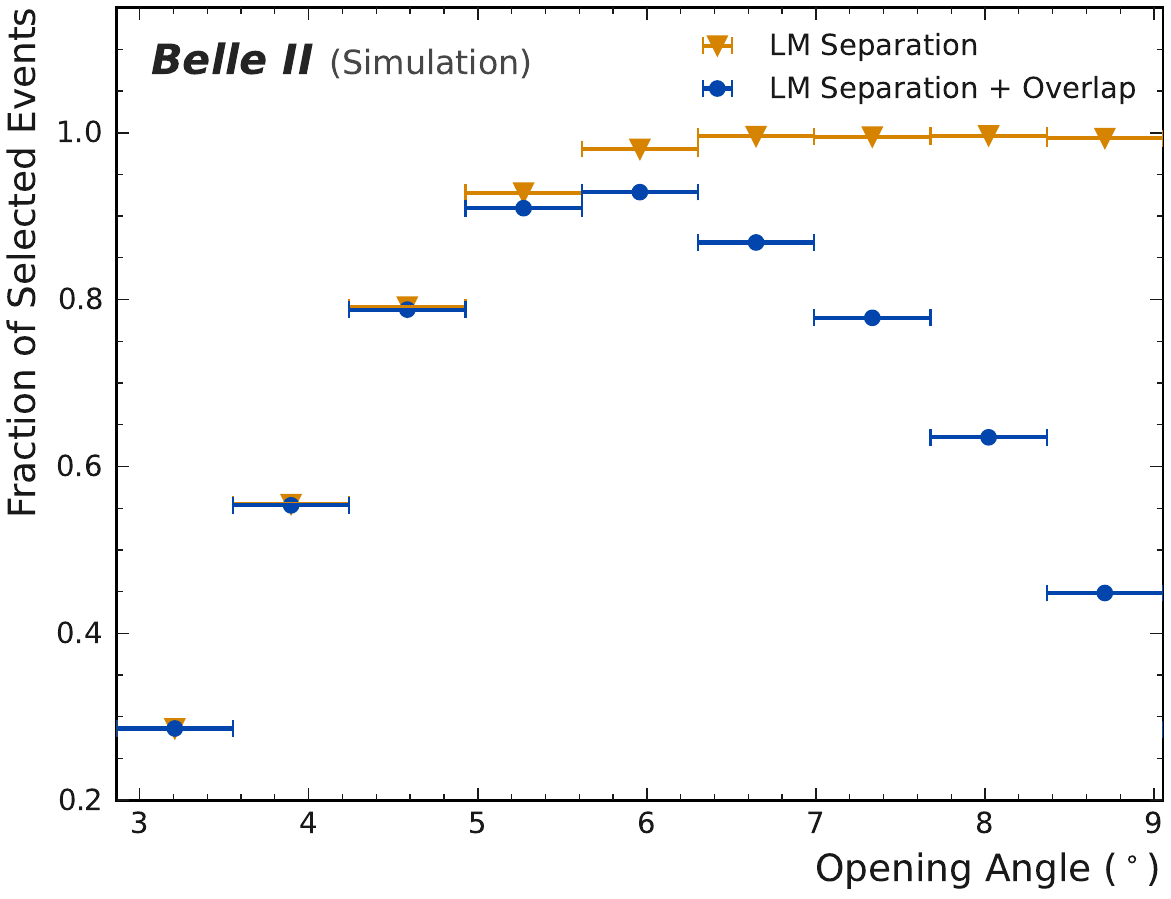}
    \caption{Fraction of selected overlapping photon events in the barrel as a function of generated opening angle. The orange markers correspond to events fulfilling LM \textit{separation criteria} a)-c); the blue markers correspond to events that additionally pass the \textit{overlap criterion} d) (see text for details).}
    \label{fig:opening_angles}
\end{figure}
Figure~\ref{fig:opening_angles} shows the fraction of events accepted by these selections as a function of the simulated opening angle. 
In the scope of this paper, we additionally require LMs to exclusively originate from simulated particles without additional LMs, e.g. from beam background, in the ROI, that is:
\begin{enumerate}[e)]
    \item the two LMs must be the only ones in the ROI and they must be truth-matched to the simulated photons.
\end{enumerate}
Finally, we remove rare cases of small truth energy depositions and large backgrounds, by requiring:
\begin{enumerate}[f)]
\item the crystal with the largest truth energy deposition of a photon must be within a $5\times5$ area around its corresponding LM.
\end{enumerate}
We then create a ROI centered at the midpoint between the two LMs,  calculated using the shortest distance between two LMs projected onto the surface of a sphere. The crystal closest to the midpoint is defined as the ROI center.
The LM positions for this are determined by interpreting the global LM coordinates of their associated crystals as latitude and longitude.
Figure~\ref{fig:edtwonominal}~(bottom) shows an overlapping photon event with high beam background.

The truth energy deposition per photon and the reconstructed crystal energy \ereccrystal, crystal time \treccrystal, crystal PSD information (see Sec.\,\ref{sec:ecl}), and the LM positions within the ROI are recorded for each event.
\section{Reconstruction Algorithms}
\label{sec:reconstruction}

Interactions of energetic photons in the \belletwo ECL typically deposit energy in up to $5\times 5$ crystals.
The task of the clustering reconstruction algorithms is to select a set of crystals that contains all the energy of the incoming photon, but no energy from other particles or from beam background.
Low beam background results in approximately $17\,\%$ of all crystals in the ECL having significant reconstructed energy $\ereccrystal\geq1\,$MeV; for high beam backgrounds this number is expected to increase to about $40\,\%$. 
This increase in the number of crystals to consider in the clustering, adds to the complexity of the reconstruction.

\subsection{Baseline}
\label{subsec:baseline}
The baseline algorithm is designed to provide maximum efficiency for cluster finding, contain all crystals from the incoming particle for particle identification, and select an optimal subset of the cluster crystals that provides the best energy resolution~\cite{Kou:2018nap}.
The clustering is performed in three steps.
In the first step, all crystals are grouped into a connected set of crystals, so-called connected regions starting with LMs, as defined previously.
In an iterative procedure all direct neighbors with energies above 0.5\,MeV are added to this LM, and the process is continued if any neighbor itself has energy above 10\,MeV.
Overlapping connected regions are merged into one.

In the second step, each connected region is split into clusters, one per LM.
If there is only one LM in the connected region, up to 21 crystals in a $5\times 5$ area excluding corners centered at the local maximum are grouped into a cluster.
If there is more than one LM in a connected region, the energy in each crystal of the connected region is assigned a distance-dependent weight and can be shared between different clusters.
The distance is calculated from the cluster centroid to each crystal center, where the cluster centroid is updated iteratively using logarithmic energy weights.
This process is repeated until all cluster centroids in a connected region are stable within 1\,mm.

In a third step, an optimal subset, including the $n$ highest energetic crystals of all non-zero weighted crystals that minimize the energy resolution, is used to predict the cluster energy \erecbasf.
$n$ depends on the measured noise in the event, and on the energy of the LM itself.
The noise level is estimated by counting the number of crystals in the event containing more than 5\,MeV that have times $t$ more than 125\,ns from the trigger time.
\erecbasf is also corrected already within \basf for possible bias using simulated events.
This bias includes leakage (energy not deposited in the crystals included in the energy sum) and beam backgrounds (energy included in the sum that is not from the signal photon).
\erecbasf is the estimator for the generated energy of a particle.

%%%%%% DEFINE THIS HERE
The \basf clustering algorithm also returns a cluster energy \erecrawbasf that is not corrected for energy bias.
\erecrawbasf is the estimator for the deposited energy of a particle.

\subsection{Graph Neural Network Architecture}
\label{subsec:gnn}
GNN architectures have shown that they are powerful network types to deal with both irregular geometries and varying input sizes. 
In this work, all crystals of an ROI with an energy deposition above 1~MeV are interpreted as nodes in a graph, which leads to variable input sizes and is thus a good use case for GNNs. The implementation of this GNN is done in \pyg~\cite{Fey/Lenssen/2019}.

The input features consist of crystal properties and crystal measurements: The global coordinates $\theta$ and $\phi$ of each crystal, the local coordinates $\theta^\prime$ and $\phi^\prime$ with respect to the ROI center, the crystal mass, and the LM(s) (in one-hot encoding) represent crystal properties. The crystal energy \ereccrystal in GeV, the time \treccrystal in $\mu$s, and the PSD fit type, PSD $\chi^2$, and PSD hadronic energy ratio are crystal measurements used as input features.
Pre-processing scales the input uniformly before further processing with the GNN: All features are min-max normalized to an interval of $[0,1]$ with the exception of \treccrystal and the PSD hadronic energy ratio which are both normalized to the interval $[-1,1]$. 
The global coordinates and the crystal masses are normalized based on the range of coordinates and masses of all crystals in the detector instead of only the ones in the ROI. % We employ min-max normalization by \texttt{scikit-learn}~\cite{scikit} for the pre-processing.
Additionally, we average each input feature over all nodes in the ROI and concatenate the averaged input features as additional inputs, thus enabling a global exchange of information.

\begin{figure}
\centering 
    \includegraphics[width=\halfwidth\textwidth]{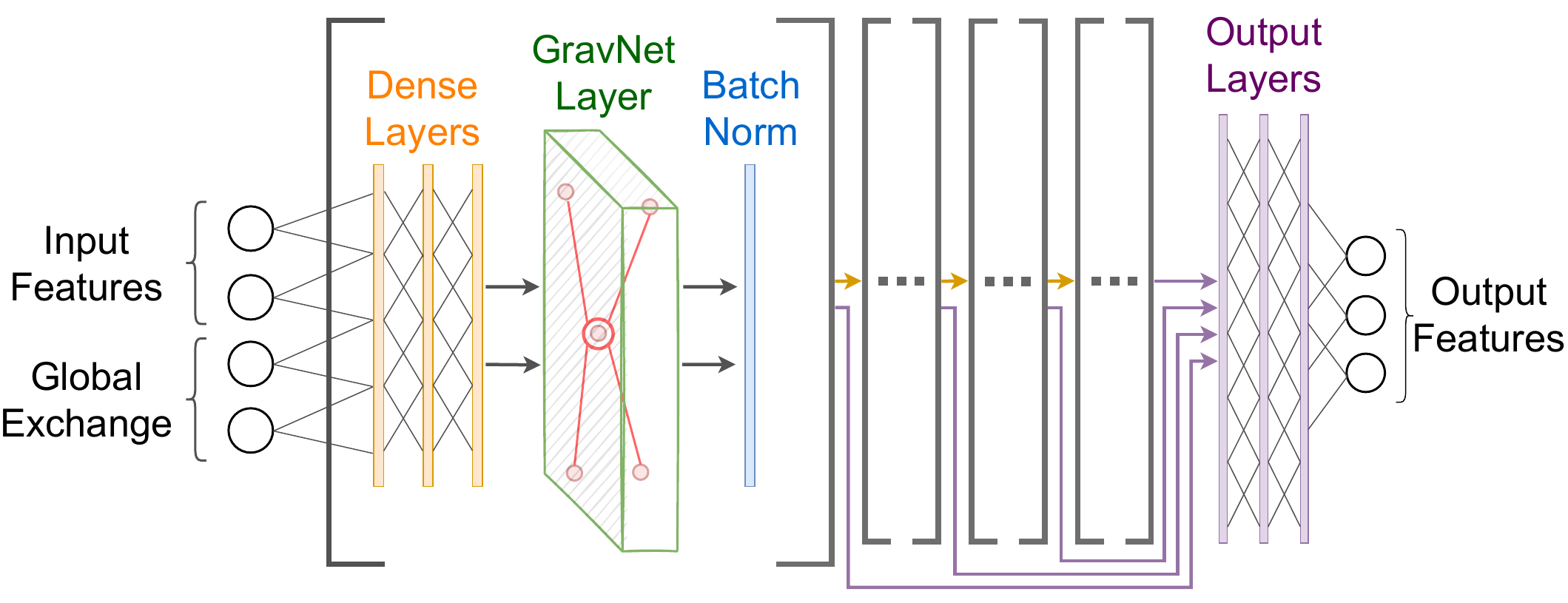}
    \caption{An illustration of the GNN architecture. Each pair of gray, square brackets represents one \gravnet block consisting of dense layers, a \gravnet layer and a batch norm layer. 
    The input features describe the feature vector of one node.
    The global exchange denotes appending the average each input features over all nodes in the ROI.}
    \label{fig:gravnet_concept}
\end{figure}

As displayed in Fig.~\ref{fig:gravnet_concept}, our model is built out of four so-called \gravnet~\cite{Qasim:2019otl} blocks of which the concatenated outputs are passed through three dense output layers with a final softmax activation function.
Each \gravnet block features three dense layers at the beginning of the block, the initial two of which with ELU~\cite{DBLP:journals/corr/ClevertUH15} activation functions and the last one with a $\tanh$ activation function. 
The dense layers feed into a \gravnet layer and the overall \gravnet block is concluded by a batch normalization layer~\cite{pmlr-v37-ioffe15}. 
The \gravnet layer is responsible for the graph building and subsequent message passing between the nodes of the graph. 
It first translates the input features into two learned representation spaces: one representing spatial information $S$ while the other, denoted $F_\mathrm{LR}$, contains the transformed features used for message passing. 
In the second step, each node is connected to its $k$ nearest neighbors defined by the Euclidean distances in $S$, thus creating an undirected, connected graph. 
For each node, the input features of connected nodes are then weighted by a Gaussian potential depending on the distance in $S$ and aggregated by summation. 
The resulting features are concatenated with the \gravnet input features and, after batch normalization, passed to the next \gravnet block and to the dense output layers.
% \add{The input to the GNN are all input features of the node as well as the the global exchange features that are the same for all nodes of the event. The actual graph per node is constructed in the \gravnet layers and has access to other nodes of the ROI.}

The implementation in the present work follows the concept of fuzzy clustering which refers to the partial assignment of individual crystals to several clustering classes. Consequently, the GNN predicts weights $w_i^\mathrm{X}$ that indicate the proportion of the reconstructed energy $E^\mathrm{crystal_i}_\mathrm{rec}$ in a crystal $i$ that belongs to a clustering class~X. For models used with isolated photons, \mbox{$\mathrm{X}\in\{\gamma_{1},\mathrm{background}\}$}, for models with overlapping photons \mbox{$\mathrm{X}\in\{\gamma_{1},\gamma_{2},\mathrm{background}\}$}. As a loss function, we then use the Mean Squared Error (MSE) between the true and predicted weights summed over all classes and crystals. The training is stopped when there has been no improvement for 15 epochs in the optimization objective. For low beam background models that objective is the MSE loss on the validation data set, whereas the high beam background models employ the more high-level FWHM$_\mathrm{dep}$ (Sec.~\ref{sec:metrics}) on the validation data set.

\begin{table*}[t!]
\centering
\caption{Optimized hyperparameters of the isolated photon, and overlapping photon \gravnet models. The hyperparameters are the result of an optimization of the FWHM$_\mathrm{dep}$ on the respective high background validation data set. }
\begin{tabular}{lrr}
\toprule
Hyperparameter                          & Isolated Photon Models & Overlapping Photon Models \\ 
\midrule    
Width of the Dense Layers, F$_\mathrm{IN},\,$F$_\mathrm{OUT}$     &         22          &          24       \\
Feature Space Dimension F$_\mathrm{LR}$         &         16          &          16        \\
Spatial Information Space Dimension S       &          6          &          6         \\
Connected Nearest Neighbors $k$                   &          14         &          16         \\
Batch Norm Momentum                 &           0.01      &          0.4         \\
Stacked GravNet Blocks                          &        4            &          4           \\
Batch Size                                      &          1024       &          512          \\
\bottomrule
\end{tabular}
\label{tab:hypparams}
\end{table*}

\begin{figure*}[t!]
     \centering
     \begin{subfigure}[b]{\third\textwidth}
         \centering
         \includegraphics[width=\textwidth]{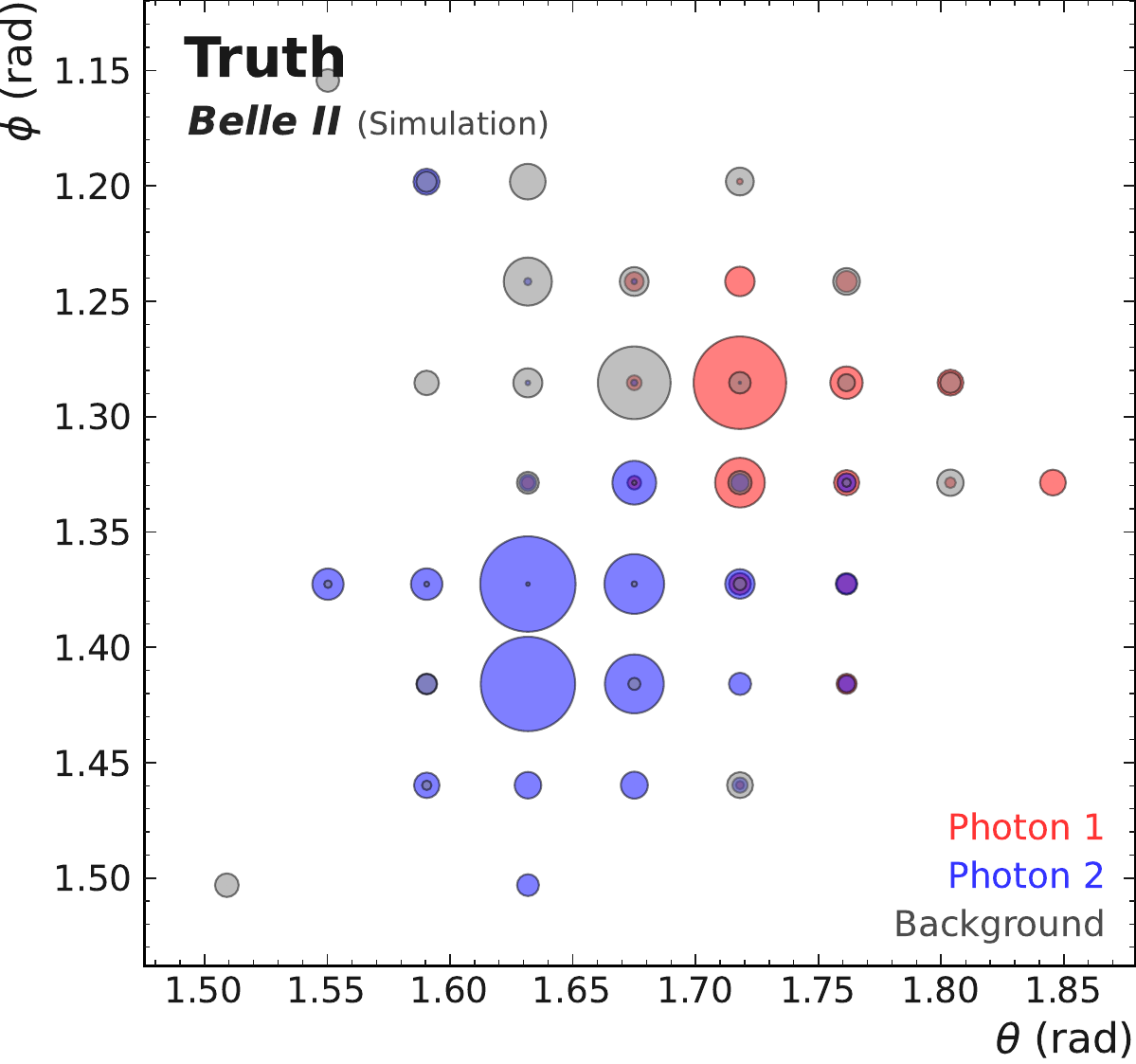}
         \caption{Truth}
         \label{fig:truth_clustering}
     \end{subfigure}
     \hfill
     \begin{subfigure}[b]{\third\textwidth}
         \centering
         \includegraphics[width=\textwidth]{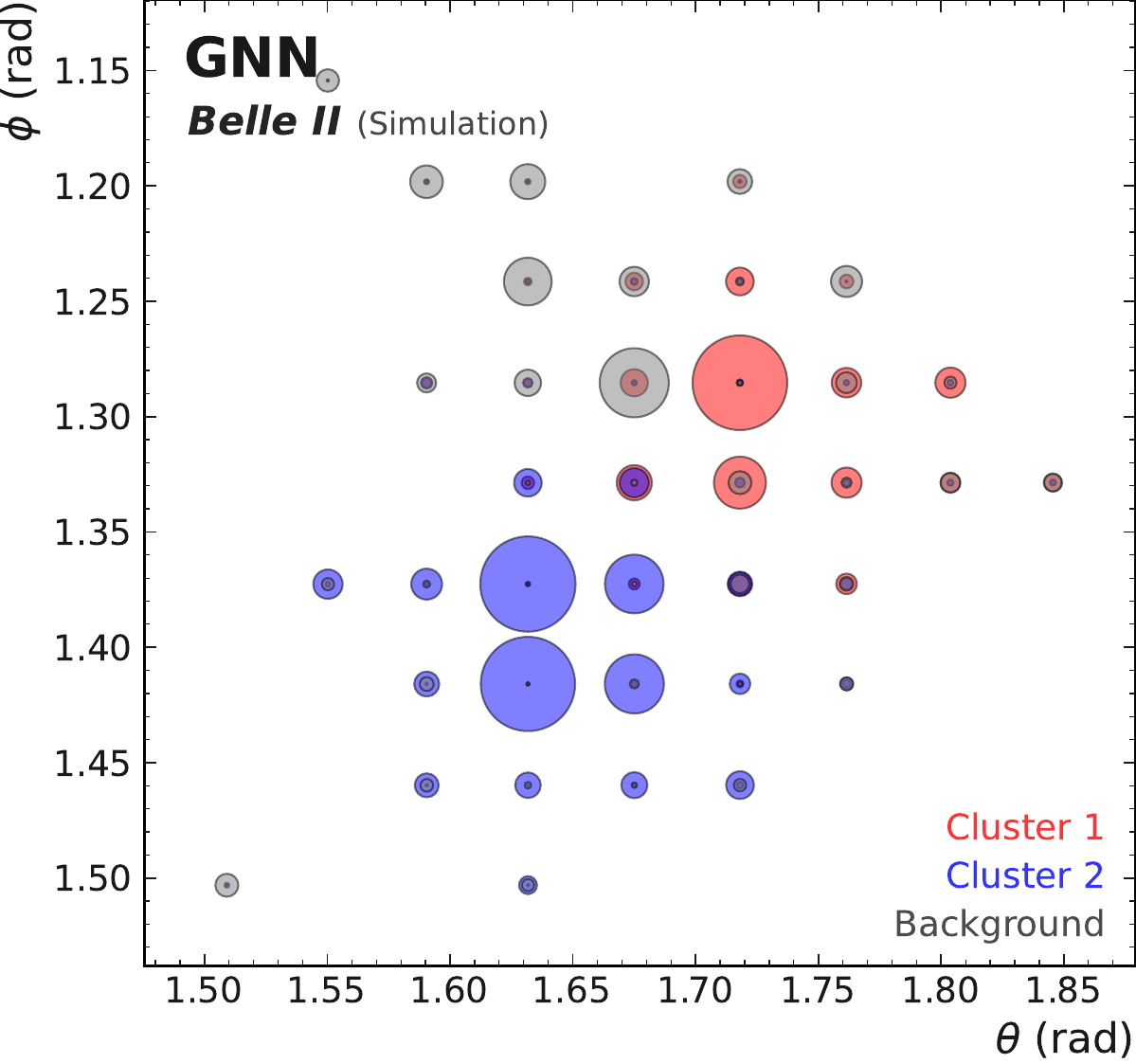}
         \caption{GNN}
         \label{fig:gnn_clustering}
     \end{subfigure}
     \hfill
     \begin{subfigure}[b]{\third\textwidth}
         \centering
         \includegraphics[width=\textwidth]{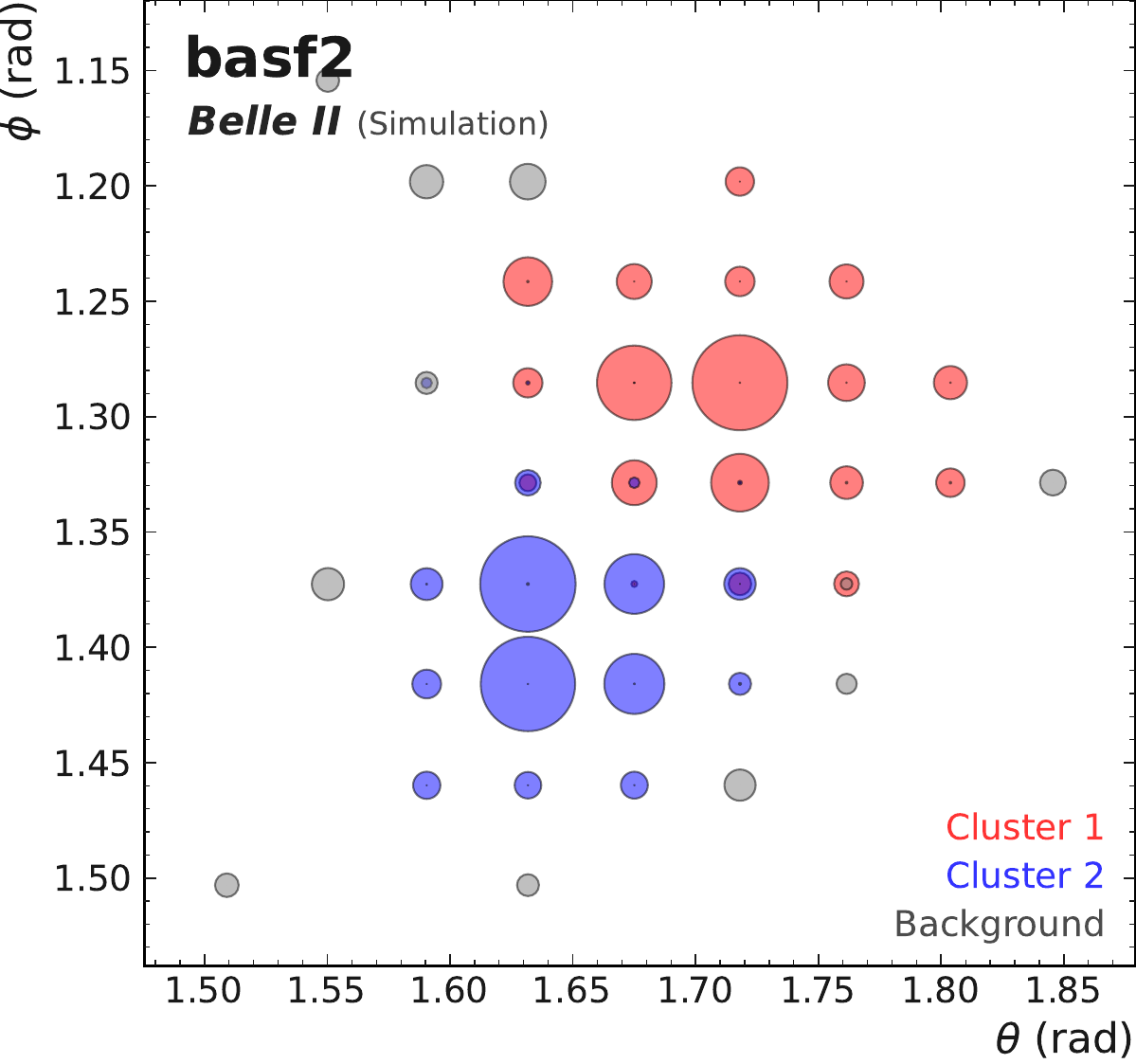}
         \caption{\basf}
         \label{fig:basf2_clustering}
     \end{subfigure}
        \caption{Comparison of (\subref{fig:truth_clustering})~truth energy fractions~, (\subref{fig:gnn_clustering})~reconstructed energy fraction by the GNN~, and (\subref{fig:basf2_clustering})~reconstructed energy fraction by \basf for an example event with high beam background. Colors indicate the fractions belonging to each photon or background. The marker centers indicate the crystal centers, the marker area is proportional to the truth or reconstructed (GNN, \basf) energy deposition respectively.}
        \label{fig:example_overlapping_cluster}
\end{figure*}

Hyperparameters have been chosen through a hyperparameter optimization using \texttt{Optuna}~\cite{optuna_2019}. 
The optimization is done with respect to the FWHM$_\mathrm{dep}$ (Sec.~\ref{sec:metrics}) instead of the loss function.
We optimize the two models trained for high beam backgrounds and use the respective hyperparameters also for the corresponding low beam background models.
The final hyperparameters for both the isolated photon models and the overlapping photon models are shown in Table \ref{tab:hypparams}.

The learning rate, the number of dense layers in each \gravnet block, and all dimensions of the output layers have been manually optimized by testing a reasonable range of values.
The learning rate is set to 5\,$\times \, 10^{-3}$ and is subject to a decay factor of 0.25 after every five epochs of stagnating validation loss. 
We did not observe significant over-training and as a consequence, we do not use dropout layers or other regularization methods but rely on the large data set.

The GNN algorithm yields the weights $w_i^\mathrm{X}$ per crystal for all crystals in the ROI with an energy deposition above 1~MeV. In order to reconstruct the total cluster energy \erecgnn associated with a certain particle, we then sum over all specific weights multiplied by the reconstructed energies per crystal, \hbox{$\erecgnn=\sum w_i^\mathrm{X}E^\mathrm{crystal_i}_\mathrm{rec}$}.

Figure \ref{fig:example_overlapping_cluster} shows how the GNN and the \basf algorithms behave in clustering a typical case of overlapping photons.
\section{Metrics}
\label{sec:metrics}
For performance evaluation, the reconstructed energy of a particle is compared with two different truth targets:   
the total deposited truth energy \edep per photon in the ROI, and the generated truth energy \egen per photon. 
This results in two variants of relative reconstruction errors.
The reconstruction error on the deposited energy 
\begin{align}\label{eq:etadep}
     \etadep^{\text{basf2}} & = \frac{\erecrawbasf-\edep}{\edep}\quad \text{and}\nonumber\\
     \etadep^{\text{GNN}} & = \frac{\erecgnn-\edep}{\edep}
\end{align}
gives access to the energy resolution ignoring leakage and other detector effects. 
It is a direct evaluation of the clustering performance of an algorithm.

On the other hand, the reconstruction error on the generated energy
\begin{align}\label{eq:etagen}
     \etagen^{\text{basf2}} & = \frac{\erecbasf-\egen}{\egen}\quad \text{and}\nonumber\\
     \etagen^{\text{GNN}} & = \frac{\erecgnn-\egen}{\egen}
\end{align}
factors in all detector and physics effects and quantifies how much of the improvements to the underlying clustering carry over to downstream physics object reconstruction.

\begin{figure}[t!]
\centering 
    \includegraphics[width=\halfwidth\textwidth]{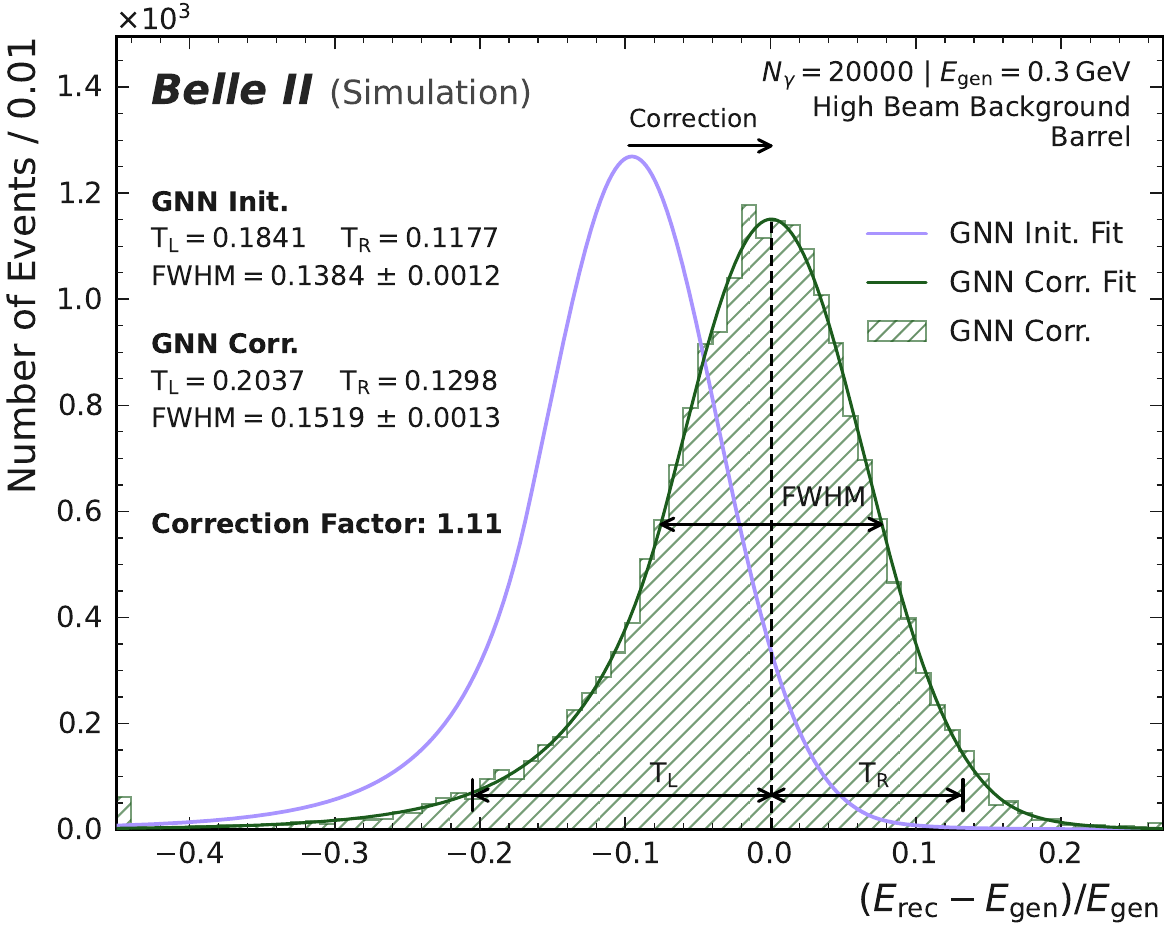}
    \caption{Example distribution of the relative reconstruction error \etagen of the generated energy and illustration of the bias correction, the \fwhm, and the tail ranges.}
    \label{fig:bias_correction}
\end{figure}

Evaluating both algorithms on a large number of simulated photons yields peaking distributions in both reconstruction errors \etadep and \etagen. 
Both distributions are potentially biased because of energy leakage and the presence of beam backgrounds (see Sec.\,\ref{subsec:baseline}).
We perform a binned fit using a double-sided crystal ball~\cite{Gaiser:Phd, Skwarnicki:1986xj} function as probability density function (pdf) with the \texttt{kafe2}~\cite{gäßler2022kafe2} framework.
We shift all reconstruction error distributions independently by a multiplicative factor to correct the difference between the fitted peak position and zero (Fig.~\ref{fig:bias_correction}). 
Since \etadep and \etagen are asymmetric distributions, we repeat this procedure until the difference between the fitted peak position and zero is less than 0.002. 
This procedure usually converges within two or three iterations.

We then determine the full width half maximum~(\fwhm) of the final shifted distributions in \etadep and \etagen, yielding \fwhmdep and \fwhmgen respectively. The uncertainty on the \fwhm is calculated from the uncertainties of the fit parameters. 
In addition to the \fwhm, we determine the tails of the reconstruction error distribution. The left and right tails $T_\text{L,R}$ are calculated as the 95th percentile when ranking the unbinned events on the respective side of the peak position, as given by the fit parameters, in ascending order ($T_\text{R}$) and descending order ($T_\text{L}$) respectively. Propagating the uncertainty on the peak position as given by the fit yields the uncertainty on $T_\text{L,R}$.
\section{Results}\label{sec:results}
\begin{figure*}[t!]
     \centering
     \begin{subfigure}[b]{\half\textwidth}
         \centering
         \includegraphics[width=\textwidth]{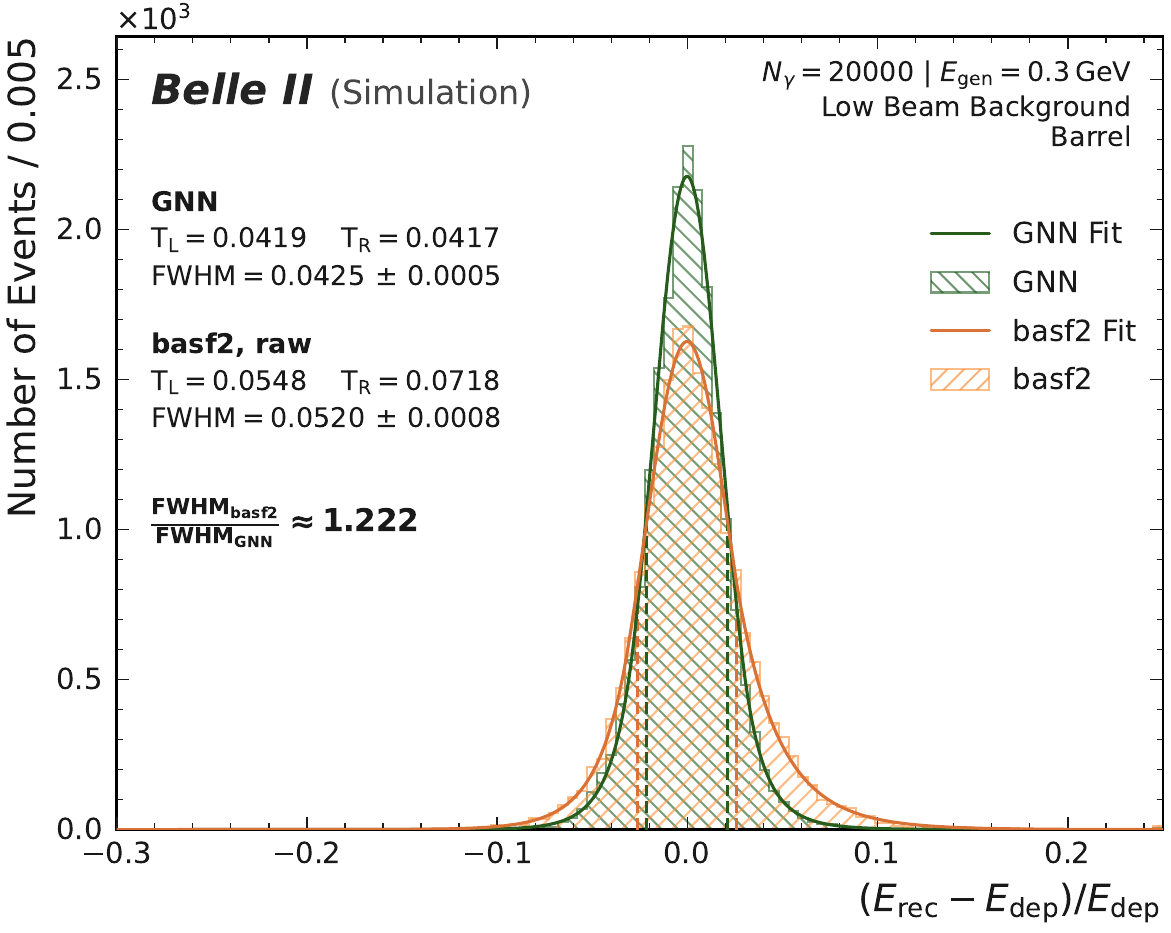}
         \caption{Relative reconstruction error \etadep of the deposited energy.}
         \label{fig:res_one_dep}
     \end{subfigure}
     \hfill
     \begin{subfigure}[b]{\half\textwidth}
         \centering
         \includegraphics[width=\textwidth]{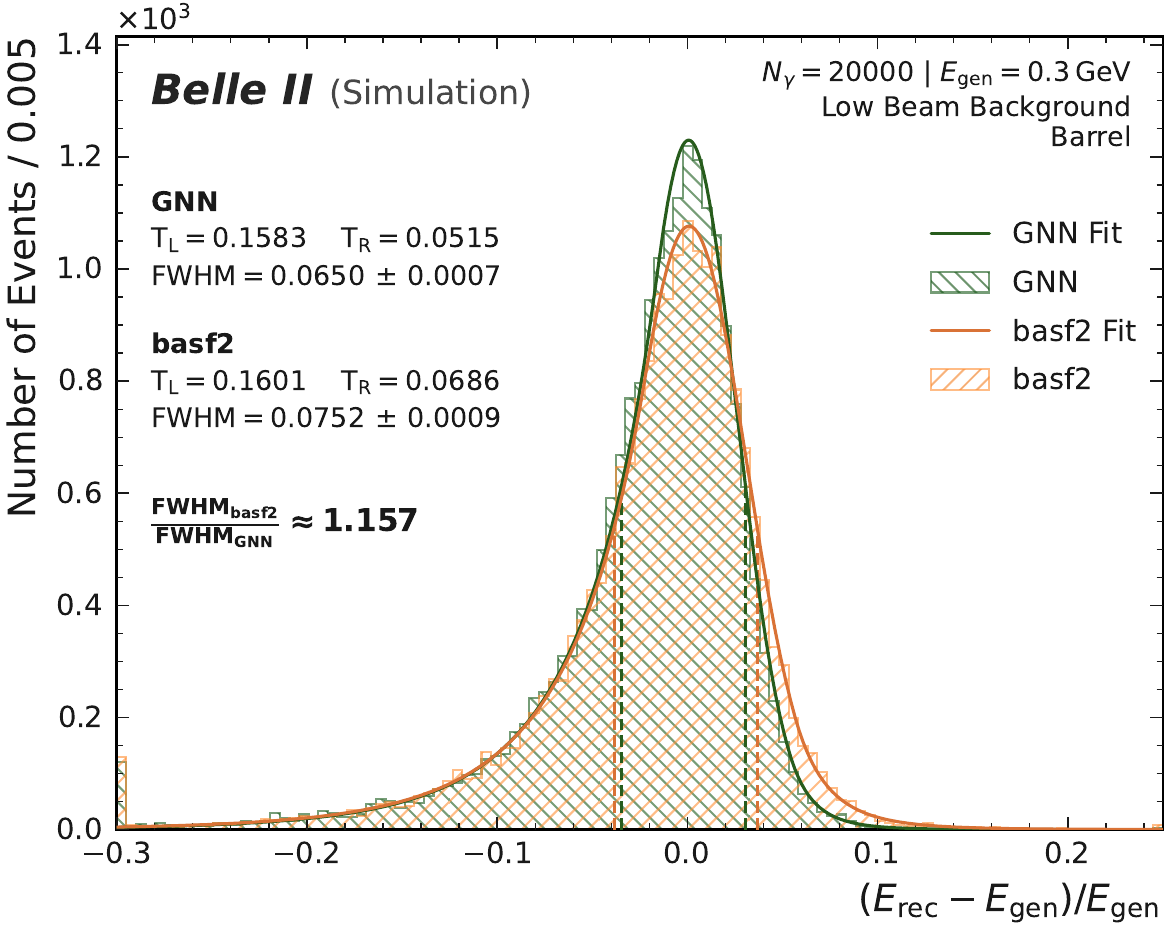}
        \caption{Relative reconstruction error  \etagen of the generated energy.}
         \label{fig:res_one_gen}
     \end{subfigure}
     
        \caption{Distribution of relative reconstruction errors (\subref{fig:res_one_dep})~\etadep and (\subref{fig:res_one_gen})~\etagen for isolated clusters for low beam backgrounds. The first bin contains all underflow entries; the last bin contains all overflow entries.}
        \label{fig:res_one}
\end{figure*}

\begin{figure*}[t!]
     \centering
     \begin{subfigure}[b]{\half\textwidth}
         \centering
         \includegraphics[width=\textwidth]{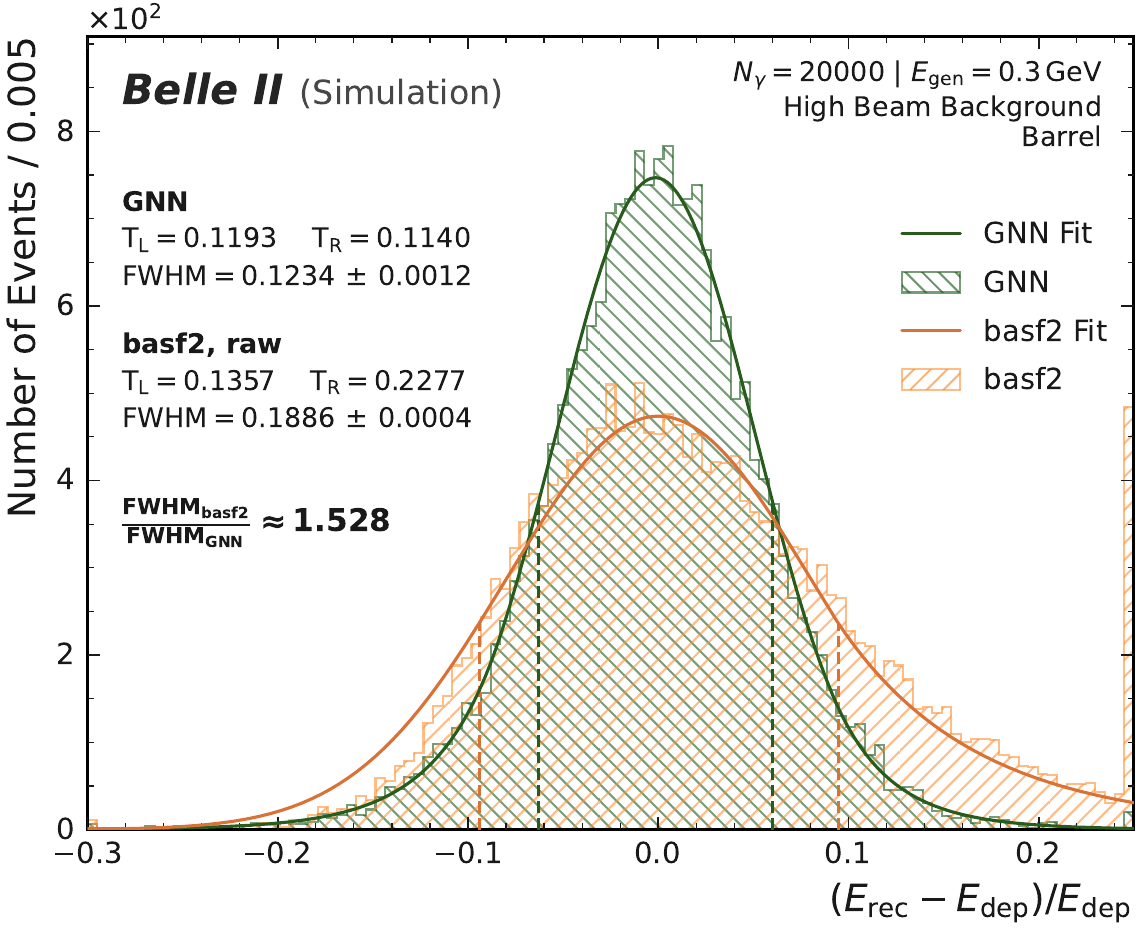}
         \caption{Relative reconstruction error \etadep of the generated energy.}
         \label{fig:res_two_dep}
     \end{subfigure}
     \hfill
     \begin{subfigure}[b]{\half\textwidth}
         \centering
         \includegraphics[width=\textwidth]{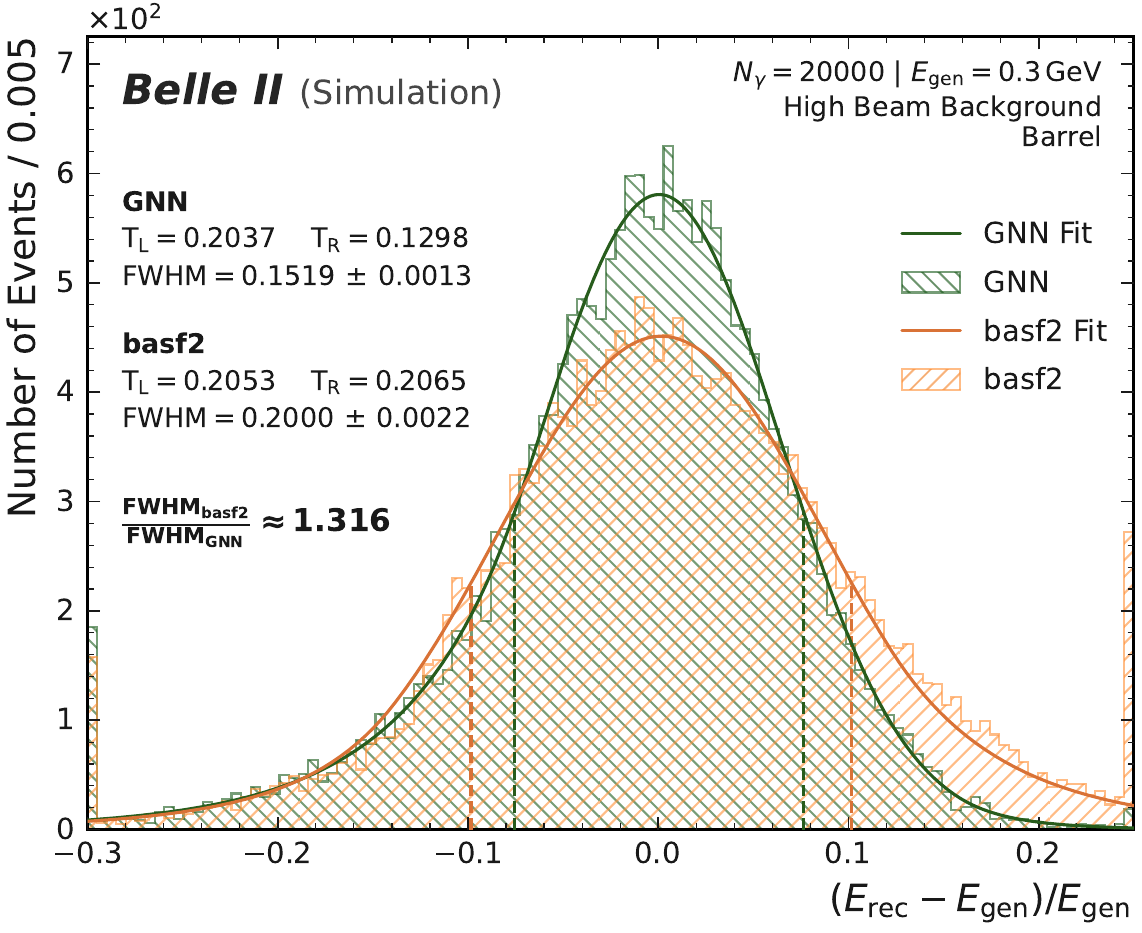}
        \caption{Relative reconstruction error \etagen of the generated energy.}
         \label{fig:res_two_gen}
     \end{subfigure}
        \caption{Distributions of relative reconstruction errors (\subref{fig:res_two_dep})~\etadep and (\subref{fig:res_two_gen}) \etagen~ for isolated clusters for high beam backgrounds. The first bin contains all underflow entries; the last bin contains all overflow entries.}
        \label{fig:res_two}
\end{figure*}

The first sections of the results focus on detailed studies of isolated clusters. Section~\ref{subsec:results_sub4} then introduces overlapping clusters and their effects on the performance. 
Figure~\ref{fig:res_one} shows examples for the distributions of both reconstruction errors $\eta_\text{dep}$ and $\eta_\text{gen}$, as well as the fit results for events with low beam background. 
Figure~\ref{fig:res_two} shows the equivalent distributions for events with high beam background.

The $\eta_\text{gen}$ distributions are wider because the reconstruction error includes the effects of leakage which result in missing energy with respect to the generated photon energy. 
This only affects the left-side tails.

In the following subsections, we are comparing the performance of the GNN and the \basf reconstruction algorithms for different detector regions for low and high beam backgrounds by evaluating the energy resolution \res and the tail parameters.
We then analyze the GNN in more detail by testing the input variable dependencies and the robustness against differences in beam background levels between training and evaluation.

\subsection{Energy resolution and energy tails}\label{subsec:results_sub3}
The three detector regions barrel, forward endcap, and backward endcap described in Sec.~\ref{sec:ecl} differ in crystal geometry, levels of background, and amount of passive material before and in between crystals. The following section studies the variations in the energy reconstruction performance that arise as a direct result of these differences.

\begin{figure*}[ht!]
     \centering
     \begin{subfigure}[b]{0.7\textwidth}
         \centering
         \includegraphics[width=\textwidth]{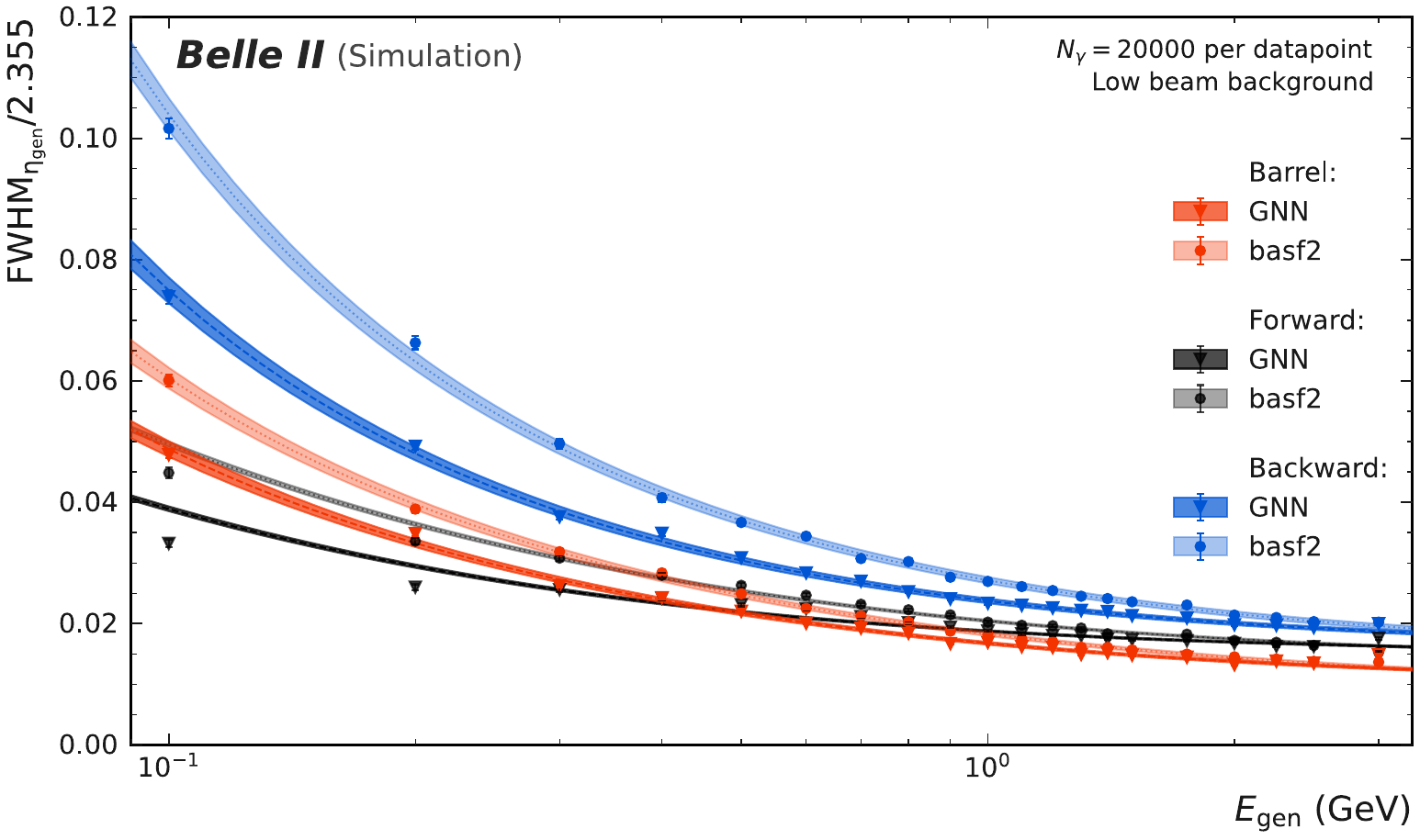}
         \caption{Low beam background.}
         \label{fig:resolution_low}
     \end{subfigure}\\
     \hfill
     
     \begin{subfigure}[b]{0.7\textwidth}
         \centering
         \includegraphics[width=\textwidth]{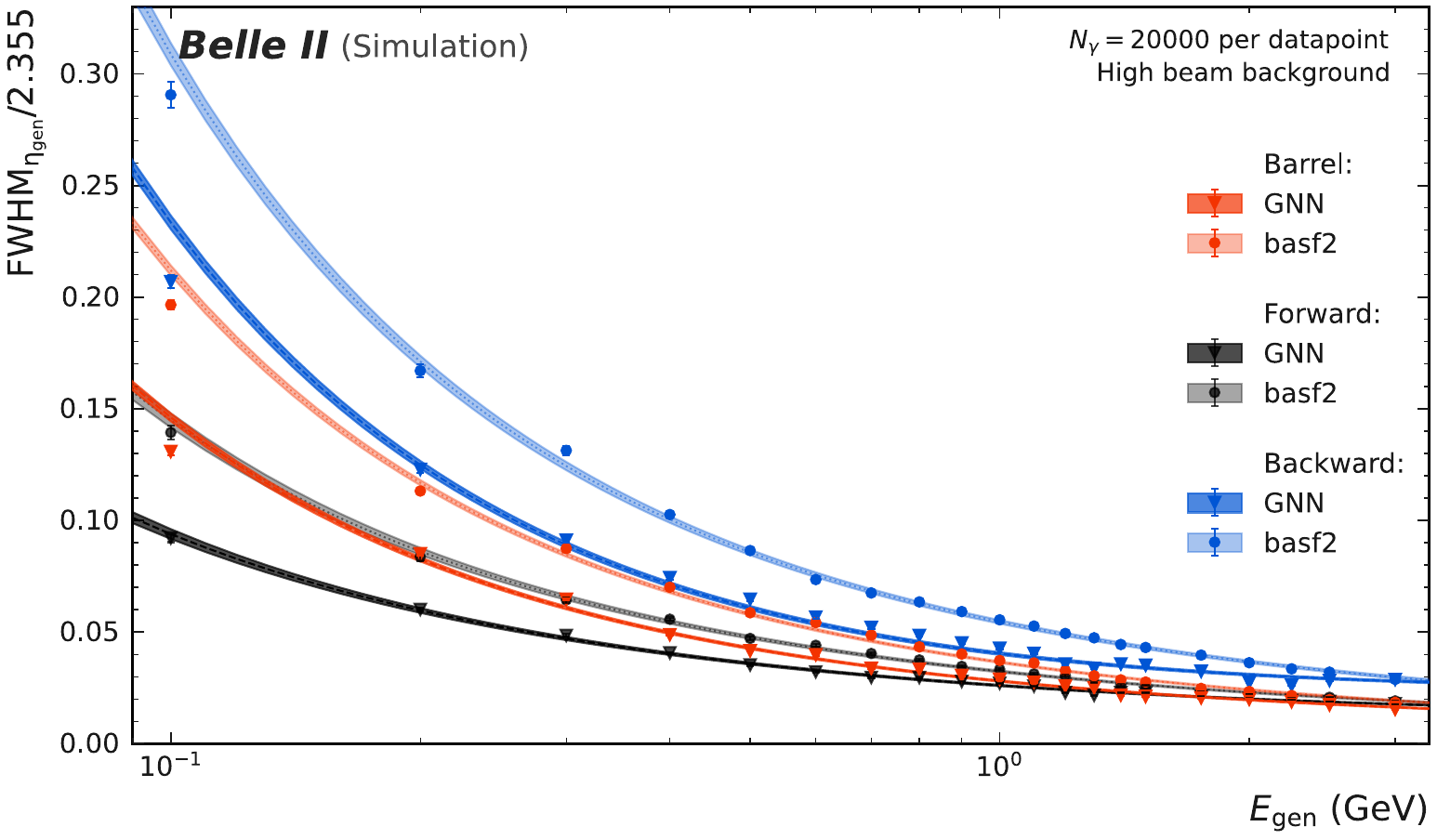}
        \caption{High beam background.}
         \label{fig:resolution_high}
     \end{subfigure}
     
        \caption{Resolution \res of the GNN and \basf as function of the simulated photon energy \egen for both endcaps and the barrel for (\subref{fig:resolution_low})~low  and (\subref{fig:resolution_high})~high  beam background. Each color is associated with one detector region; the light color indicates \basf, the dark color the GNN. The bands indicate the uncertainty of the fits, see text for details. The fit parameters are summarized in Tab.\,\ref{tab:fit_params_barrel_endcaps}.}
        \label{fig:resolution}
\end{figure*}

In order to access the energy dependence of the resolution and tail parameters we simulate test data sets of photons at various fixed energies. The \fwhm for each simulated data set is then determined according to Sec.~\ref{sec:metrics}. Plotting the resolutions \res over the generated photon energies $E_\mathrm{gen}$ reveals a characteristic relationship that is parameterized by the function $a / E_\mathrm{gen} \oplus b / \sqrt{E_\mathrm{gen}} \oplus c$, where $\oplus$ indicates addition in quadrature.

Both the GNN as well as the baseline algorithm perform differently in regards to the energy resolution in all three detector parts, as can be seen in Fig.~\ref{fig:resolution_low} for low beam background and as Fig.~\ref{fig:resolution_high} for high beam background. 
Table~\ref{tab:fit_params_barrel_endcaps} reports the parameters of the fitted parameterization of the resolution.
We attribute these difference to the large spread of both shape and size of crystals in the endcaps, the asymmetric distribution of beam backgrounds, and the different amount of passive material in front of the different detector regions.

Overall, the energy resolution of the GNN algorithm is significantly better than the baseline algorithm for all photon energies.
The GNN energy resolution is better by more than 30\,\% for photon energies below $500\,\mev$ which is the energy range of more than 90\,\% of all photons in $B$-meson decay chains.
The higher the beam background, the larger the difference between the GNN and the baseline algorithm.
The difference between the two algorithms decreases with energy because the relative contribution of beam backgrounds to the photon energy resolution decreases.

\begin{figure*}
     \centering
     \begin{subfigure}[b]{\half\textwidth}
         \centering
        \includegraphics[width=\textwidth]{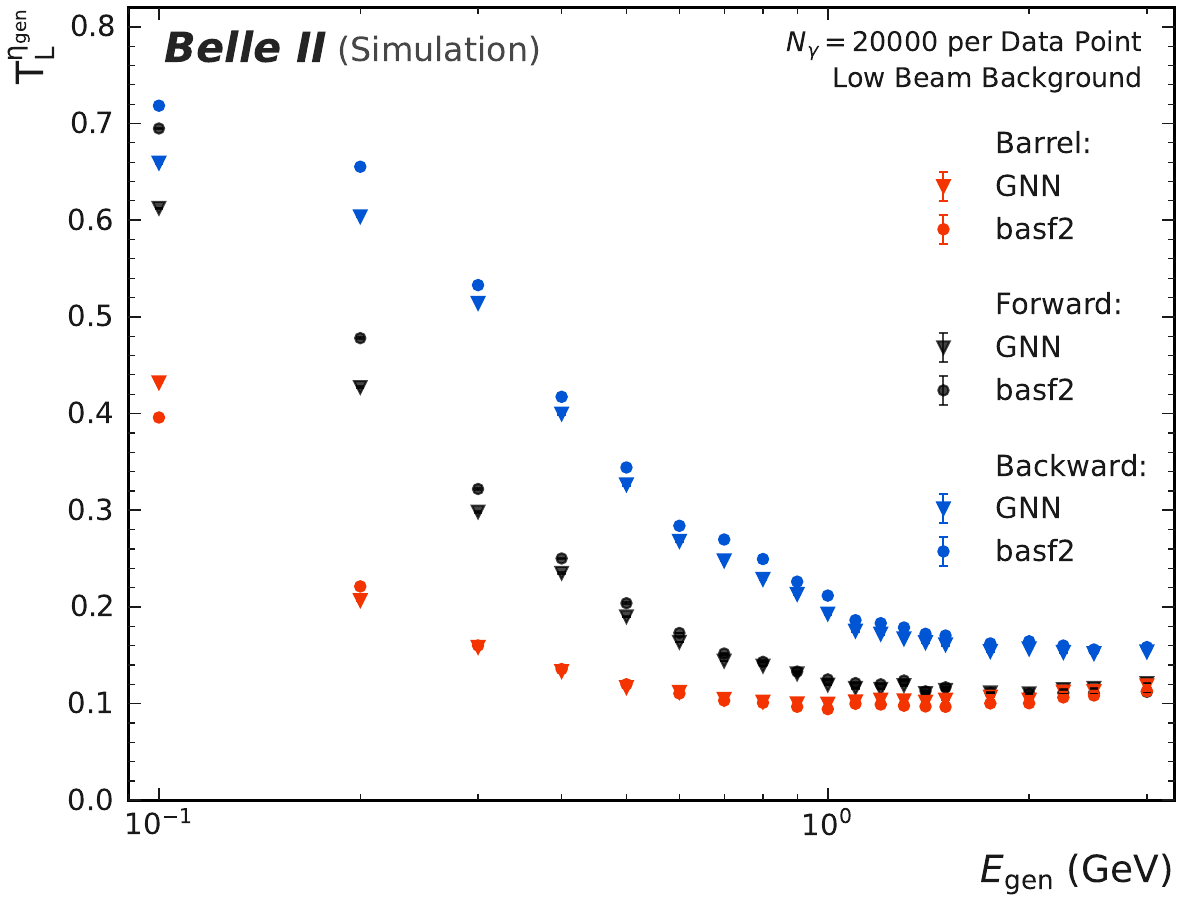}
        \caption{Left tail length $T_L$.}
        \label{fig:tails_left_early}
     \end{subfigure}
     \hfill
     \begin{subfigure}[b]{\half\textwidth}
         \centering
        \includegraphics[width=\textwidth]{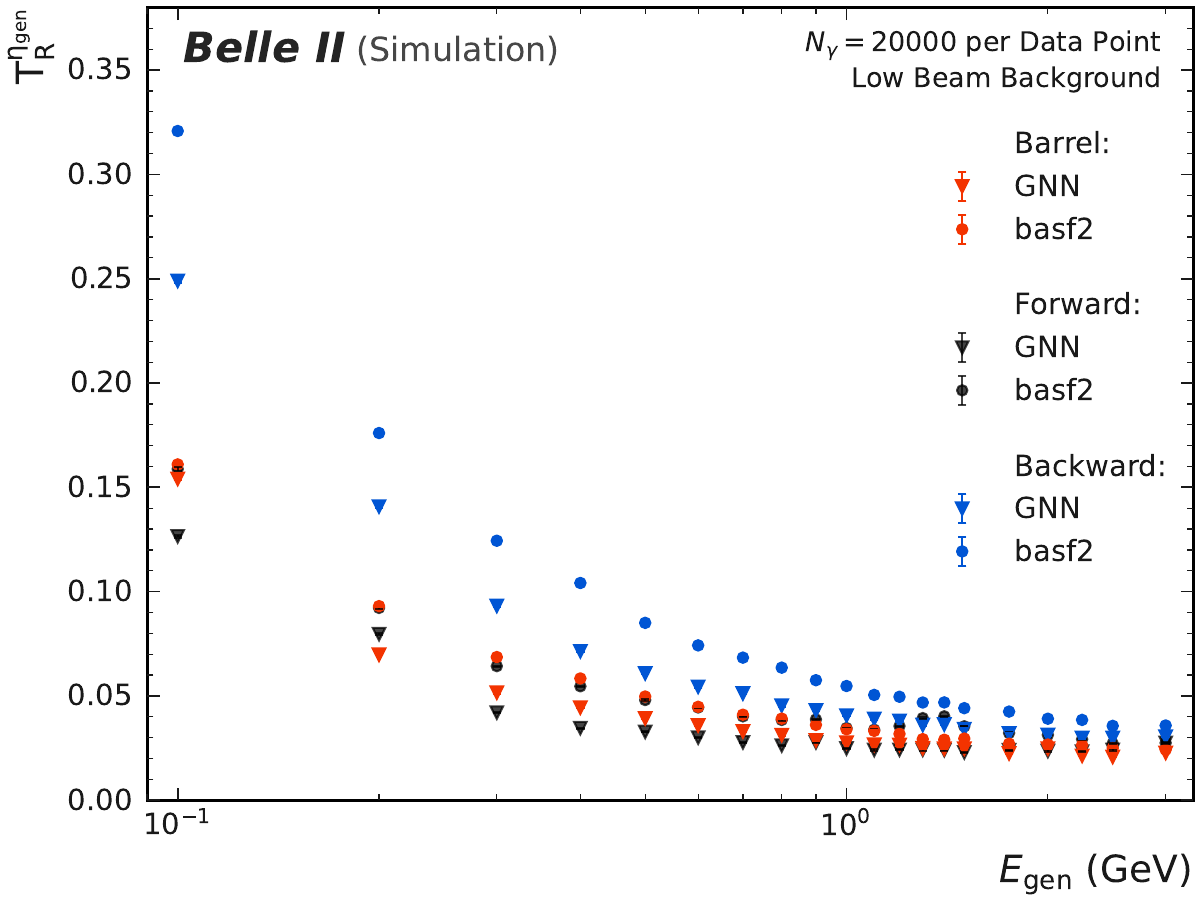}
        \caption{Right tail length $T_R$.}
        \label{fig:tails_right_early}
     \end{subfigure}
%need this empty line

     \begin{subfigure}[b]{\half\textwidth}
         \centering
        \includegraphics[width=\textwidth]{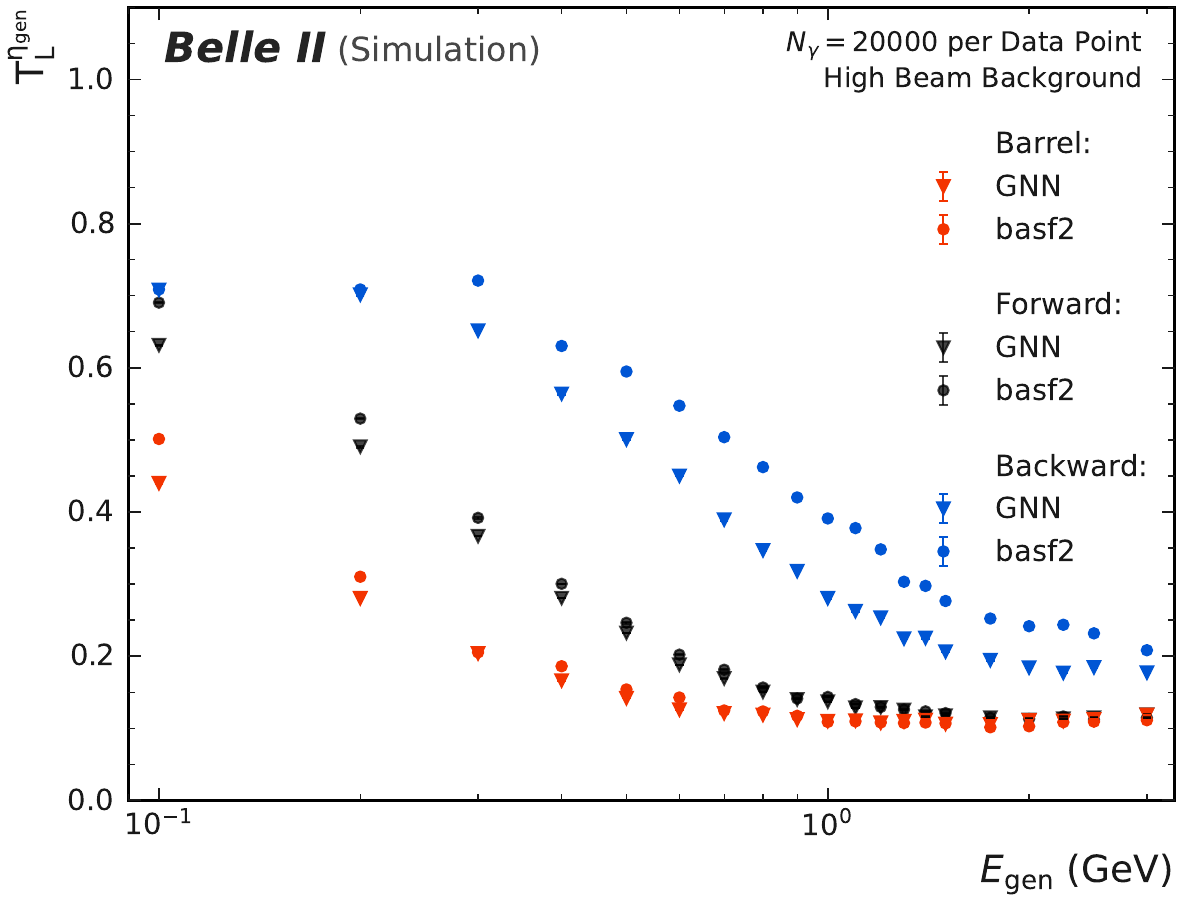}
        \caption{Left tail length $T_L$.}
        \label{fig:tails_left_nominal}
     \end{subfigure}
     \hfill
     \begin{subfigure}[b]{\half\textwidth}
         \centering
        \includegraphics[width=\textwidth]{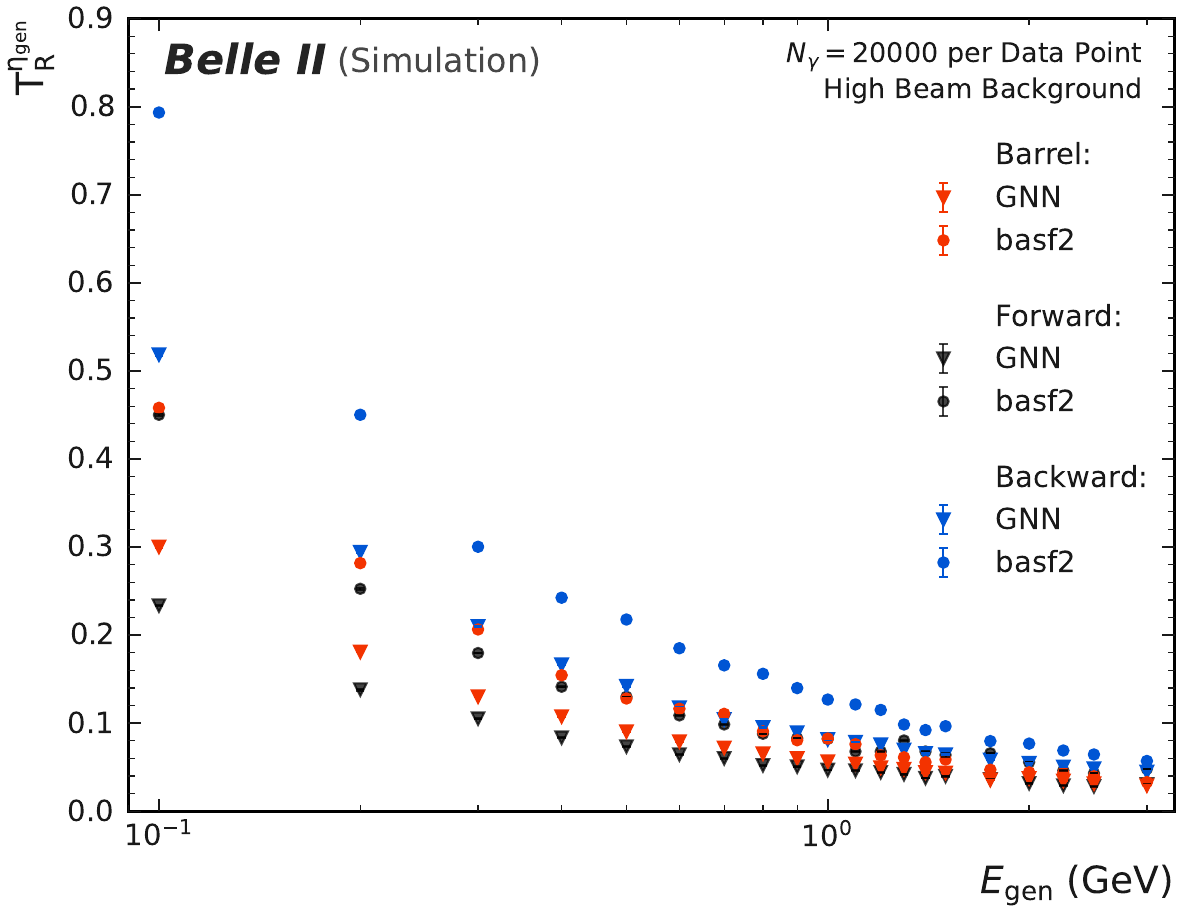}
        \caption{Right tail length $T_R$.}
        \label{fig:tails_right_nominal}
     \end{subfigure}
        \caption{95\,\% left- and right tail lengths $T_L$ and $T_R$ of \etagen for the GNN and \basf as function of the simulated photon energy \egen for both endcaps and the barrel for (\subref{fig:tails_left_early} and \subref{fig:tails_right_early})~low and (\subref{fig:tails_left_nominal} and \subref{fig:tails_right_nominal})~high beam background. Each color is associated with one detector region.}
        \label{fig:tails_nominal}
\end{figure*}

The shape of the left-side tails is dominated by passive material and is hence expected to be different in the different detector regions.
The left-side tails are almost independent of beam backgrounds as can be seen by comparing Fig.~\ref{fig:tails_left_early} for low beam background and Fig.~\ref{fig:tails_left_nominal} for high beam background. 
The GNN and the baseline algorithm both show the smallest tail length for the barrel region with decreasing tail lengths for increasing energy. 
The left-side tails are largest in the backward endcap due to the highest ratio of passive to active material as expected.
The right-side tails are mostly originating from beam background being wrongly added to photon clusters.
The GNN produces shorter tails than the baseline algorithm for all energies and for both low and high beam backgrounds, with the performance difference increasing for lower energies and higher beam backgrounds. 

\begin{table*}
\centering
\caption{Fit results ($a/E_\mathrm{gen}\oplus b / \sqrt{E_\mathrm{gen}} \oplus c$) of the fits shown in Fig.\,\ref{fig:resolution}.}
\begin{tabular}{llllllll}
\toprule
\multirow{2}{*}{\vspace{-1em}Region} & \multirow{2}{*}{\vspace{-1em}Algorithm} & \multicolumn{3}{l}{Low Beam Background} & \multicolumn{3}{l}{High Beam Background} \\
                                 &                            & a ($\times 10^{-2}$)           & b ($\times 10^{-2}$)           & c ($\times 10^{-2}$)                           &  a ($\times 10^{-2}$)           & b ($\times 10^{-2}$)            & c ($\times 10^{-2}$)                            \\
\midrule 
\multirow[t]{2}{*}{Barrel}   & GNN       & 0.23$\pm$0.02        & 1.32$\pm$0.02      & 1.00$\pm$0.01           & 1.25$\pm$0.02        & 2.39$\pm$0.02      & 0.75$\pm$0.03\\
                             & \basf     & 0.35$\pm$0.02        & 1.54$\pm$0.02      & 0.91$\pm$0.02           & 1.88$\pm$0.02        & 3.11$\pm$0.03      & 0.31$\pm$0.10\\
\multirow[t]{2}{*}{Forward}  & GNN       & 0.00$+$0.14        & 1.11$\pm$0.01      & 1.49$\pm$0.00           & 0.61$\pm$0.03        & 2.23$\pm$0.02      & 1.20$\pm$0.02\\
                             & \basf     & 0.00$+$0.37        & 1.51$\pm$0.01      & 1.38$\pm$0.01           & 1.11$\pm$0.03        & 2.92$\pm$0.03      & 0.84$\pm$0.03\\
\multirow[t]{2}{*}{Backward} & GNN       & 0.50$\pm$0.02        & 1.69$\pm$0.03      & 1.59$\pm$0.02           & 2.18$\pm$0.03       & 2.51$\pm$0.05      & 2.28$\pm$0.02\\
                             & \basf     & 0.78$\pm$0.03        & 2.12$\pm$0.04      & 1.50$\pm$0.03           & 2.72$\pm$0.05        & 4.64$\pm$0.05      & 0.91$\pm$0.08\\
\bottomrule
\end{tabular}
\label{tab:fit_params_barrel_endcaps}
\end{table*}

\subsection{Beam Background Robustness}\label{subsec:results_sub1}
\begin{figure*}[ht!]
        \centering
        \includegraphics[width=0.66\textwidth]{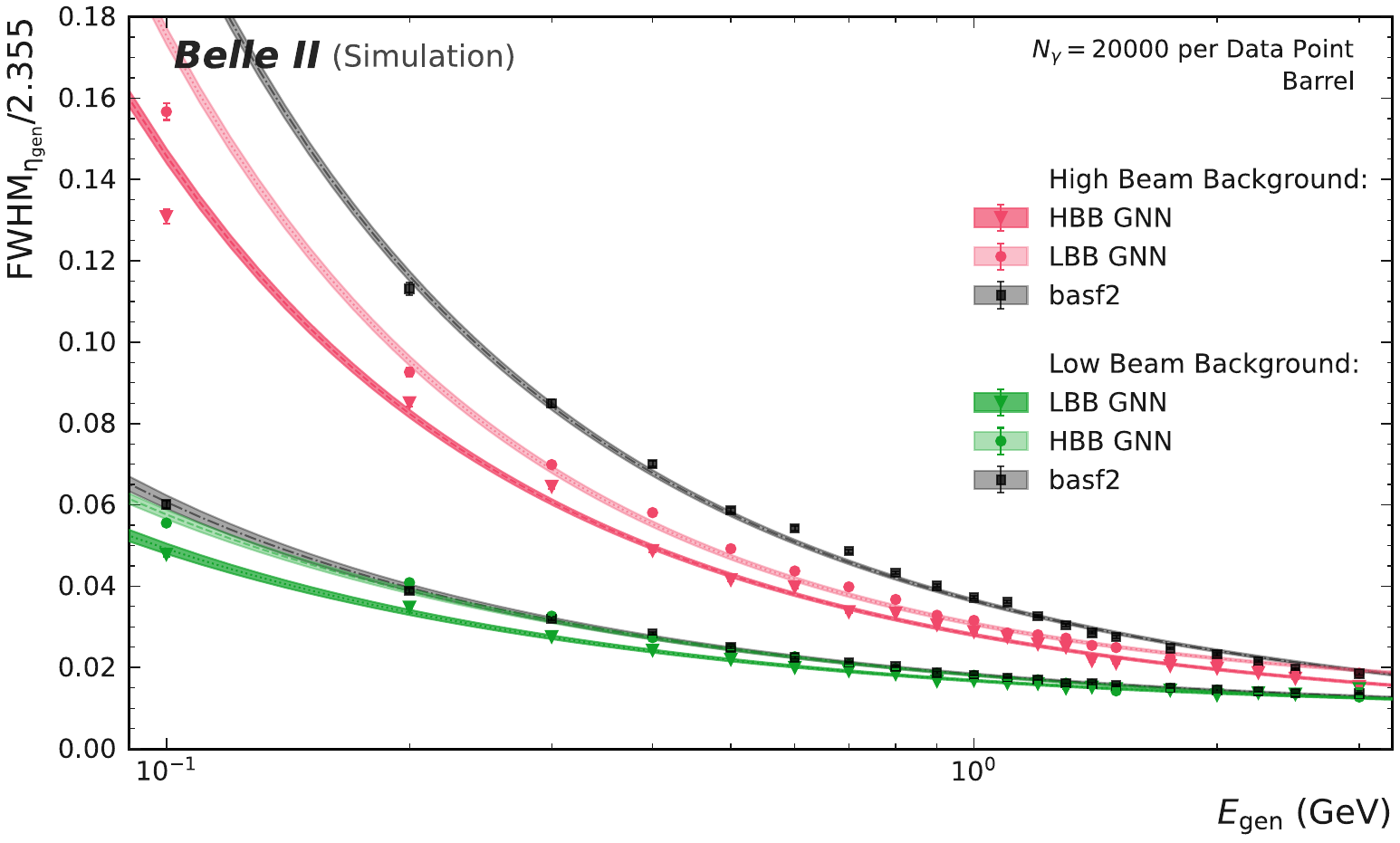}
        \caption{Resolution \res as a function of the simulated photon energy \egen for the GNNs trained with low beam background (LBB GNN) and high beam background (HBB GNN) in the barrel. The color is associated with the evaluation on either beam background; the dark color indicates the model trained with the beam background identical to the evaluation, and the light color indicates the model trained with the respective other beam background. The bands indicate the uncertainty of the fits, see text for details. The fit parameters are summarized in Tab.\,\ref{tab:fit_params_barrel_rightwrong}. The resolution of the \basf algorithm is shown for comparison.}
        \label{fig:wrong_background_barrel}
\end{figure*}

The beam background levels are changing continuously during detector operations.
Ideally, reconstruction algorithms at \belletwo are insensitive to such changes.
The \basf baseline algorithm achieves robustness against increasing beam backgrounds by adaptively including fewer crystals in the energy sum calculation.
Since our GNN is trained with a large number of events with event-by-event fluctuations of beam backgrounds, we expect robustness against varying beam backgrounds if the GNN generalizes well enough.
We test the robustness of our GNN by comparing GNNs trained and tested on the same backgrounds, against GNNs trained and tested on the two different beam backgrounds (Fig.~\ref{fig:wrong_background_barrel}, parameterization in Tab.~\ref{tab:fit_params_barrel_rightwrong}).
While the GNNs trained on the same beam backgrounds achieve a better resolution than the ones trained on different beam backgrounds, the GNN still outperforms the baseline algorithm even for networks trained on the different beam backgrounds.
This demonstrates an promising generalization with respect to different levels of beam backgrounds.

\begin{table*}[t]
\centering
\caption{Fit results ($a /E_\mathrm{gen} \oplus b / \sqrt{E_\mathrm{gen}} \oplus c$) of the fits shown in Fig.~\ref{fig:wrong_background_barrel} for the GNN trained with low beam background (LBB GNN) and high beam background (HBB GNN). The values for the LBB GNN inferred on low beam background test samples, and for the HBB GNN inferred on high beam background are identical to the ones reported in Tab.~\ref{tab:fit_params_barrel_endcaps}.}
\begin{tabular}{llllllll}
\toprule
\multirow{2}{*}{\vspace{-1em}Region} & \multirow{2}{*}{\vspace{-1em}Algorithm} & \multicolumn{3}{l}{Low Beam Background} & \multicolumn{3}{l}{High Beam Background} \\
                                 &                            & a ($\times 10^{-2}$)           & b ($\times 10^{-2}$)           & c ($\times 10^{-2}$)                           &  a ($\times 10^{-2}$)           & b ($\times 10^{-2}$)            & c ($\times 10^{-2}$)                            \\
\midrule 
\multirow[t]{2}{*}{Barrel}   & LBB GNN       & 0.23$\pm$0.02        & 1.32$\pm$0.02      & 1.00$\pm$0.01          & 1.59$\pm$0.02        & 2.27$\pm$0.03      & 1.32$\pm$0.02\\
                             & HBB GNN       & 0.28$\pm$0.02        & 1.58$\pm$0.01      & 0.85$\pm$0.02           & 1.25$\pm$0.02       & 2.39$\pm$0.02     & 0.75$\pm$0.03\\
\bottomrule
\end{tabular}
\label{tab:fit_params_barrel_rightwrong}
\end{table*}

\subsection{Input Parameter Dependency}
\label{subsec:results_sub2}

As discussed in Sec.~\ref{sec:ecl}, multiple input features are available for the GNN, while the \basf algorithm uses crystal position and energy only.
This section presents a study of the influence of the input features on the \fwhm.
For that, the architecture described in Sec.~\ref{subsec:gnn} is trained on isolated photon events with low or high beam backgrounds using different combinations of input features.
The 200\,000 events from the respective validation data set, as described in Sec.~\ref{sec:dataset}, are used for inference. The data set covers an energy range of $0.1 < \egen < 1.5\,\text{GeV}$ and the full detector range $17^{\circ} < \thetagen < 150^{\circ}$ and $0^{\circ} < \phigen < 360^{\circ}$, each of which in uniform distribution.
The \fwhm of \egen and \edep is calculated as described in Sec.~\ref{sec:metrics}.
All GNNs use the global crystal coordinates, the LM position, and the crystal mass as input features.
A comparison of the \fwhm for the different additional input features is shown in Tab.~\ref{tab:feature_ana}.
The results show, that even for the minimal set of input variables, the GNN's \fwhm is smaller than \basf's for both the deposited and the generated energy in both beam background scenarios.
Adding local coordinates leads to small improvements and using time information brings significant improvement in the GNN performance.
PSD information has almost no effect on the \fwhm.
Since the main purpose of the PSD information is to differentiate electromagnetic and hadronic interactions per crystal, this is expected.
In anticipation of future extensions of the GNN to  hadronic interactions as well, the PSD information is kept throughout this work.

\begin{table*}
\centering
\caption{Comparison of the performances of GNN models with different additional input features, and the performance of the \basf baseline. Shown are the $\mathrm{FWHM_{dep}}$ and $\mathrm{FWHM_{gen}}$ (see Sec.~\ref{sec:metrics}), for 200\,000 events in the validation data sets (see Sec.~\ref{sec:dataset}) with low and high beam background.  The data sets cover an energy range of $0.1 < \egen < 1.5\,\text{GeV}$ and the full detector range $17^{\circ} < \thetagen < 150^{\circ}$ and $0^{\circ} < \phigen < 360^{\circ}$, each of which in uniform distribution. The uncertainties of the FWHM in each column are correlated  since they use the same simulated events. The input features are described in detail in Sec.~\ref{sec:ecl}.}
\begin{tabularx}{\textwidth}{llXXXX}
\toprule
\multirow{3}{*}{\vspace{-5em}\hspace{-0.4em} Input Features} & \multicolumn{2}{l}{\multirow{2}{*}{Low Beam Background}} & \multicolumn{2}{l}{\multirow{2}{*}{\hspace{0.8em}High Beam Background}} \\
                   & \multicolumn{2}{l}{}                   & \multicolumn{2}{l}{}                   \\
                   & \makecell[c]{\makecell[l]{$\mathrm{FWHM_{dep}}$\\ $\times10^{-2}$}} & \makecell[c]{\makecell[l]{$\ \ \mathrm{FWHM_{gen}}$\\ $\ \ \times10^{-2}$}} 
                   & \makecell[c]{\makecell[l]{$\mathrm{FWHM_{dep}}$\\ $\times10^{-2}$}} & \makecell[c]{\makecell[l]{$\ \ \mathrm{FWHM_{gen}}$\\ $\ \ \times10^{-2}$}} \\
\midrule
Energy                                 & \makecell[c]{2.17$\pm$0.01} & \makecell[c]{5.25$\pm$0.02} & \makecell[c]{5.05$\pm$0.03} & \makecell[c]{8.08$\pm$0.04} \\
Energy, local coordinates              & \makecell[c]{2.11$\pm$0.02} & \makecell[c]{5.19$\pm$0.02} & \makecell[c]{5.04$\pm$0.04} & \makecell[c]{8.04$\pm$0.04} \\
Energy, local coordinates, PSD         & \makecell[c]{2.19$\pm$0.01} & \makecell[c]{5.20$\pm$0.02} & \makecell[c]{5.06$\pm$0.03} & \makecell[c]{8.07$\pm$0.04} \\
Energy, local coordinates, time        & \makecell[c]{1.72$\pm$0.01} & \makecell[c]{4.85$\pm$0.02} & \makecell[c]{4.52$\pm$0.03} & \makecell[c]{7.63$\pm$0.03} \\
Energy, local coordinates, time, PSD   & \makecell[c]{1.72$\pm$0.01} & \makecell[c]{4.85$\pm$0.02} & \makecell[c]{4.51$\pm$0.03} & \makecell[c]{7.62$\pm$0.03} \\

\midrule
\basf & \makecell[c]{2.32$\pm$0.02} & \makecell[c]{5.13$\pm$0.02} & \makecell[c]{6.73$\pm$0.05} & \makecell[c]{8.97$\pm$0.07}  \\
\bottomrule
\end{tabularx}
\label{tab:feature_ana}
\end{table*}

\subsection{Overlapping Photons}\label{subsec:results_sub4}

When discussing overlapping photon events, it is important to note that the FWHM of the photon energy distribution not only depends on its own properties but also on the properties of the second photon present.
To account for that, the evaluation is split in energy bins of [0.1, 0.2], [0.2, 0.5], [0.5, 1.0], and [1.0, 1.5]\,\gev for both photons respectively.
We report the FWHM of the first photon for different simulated energies of the second photon for low beam backgrounds (see Tab.~\ref{tab:fwhm_low}) and high beam backgrounds (Tab.~\ref{tab:fwhm_high}).

The GNN provides a better \fwhm for all combinations, but the improvement is most significant if the photon is low energetic.
For low beam backgrounds, the GNN improves the \fwhm by up to 20\,\% for photons with simulated energies between $0.1 < \egen < 0.2$\,\gev.
For high beam backgrounds, the GNN improves the \fwhm by more than 35\,\% for photons with simulated energies between $0.1 < \egen < 0.2$\,\gev.

The result shows that the significant performance improvement observed for isolated photons can also be achieved for the more complicated overlapping photon signatures.

\section{Conclusion and Outlook}
\label{sec:conclusion}
\begin{figure*}[ht!]
     \centering
     \begin{subfigure}[b]{\third\textwidth}
         \centering
         \includegraphics[width=\textwidth]{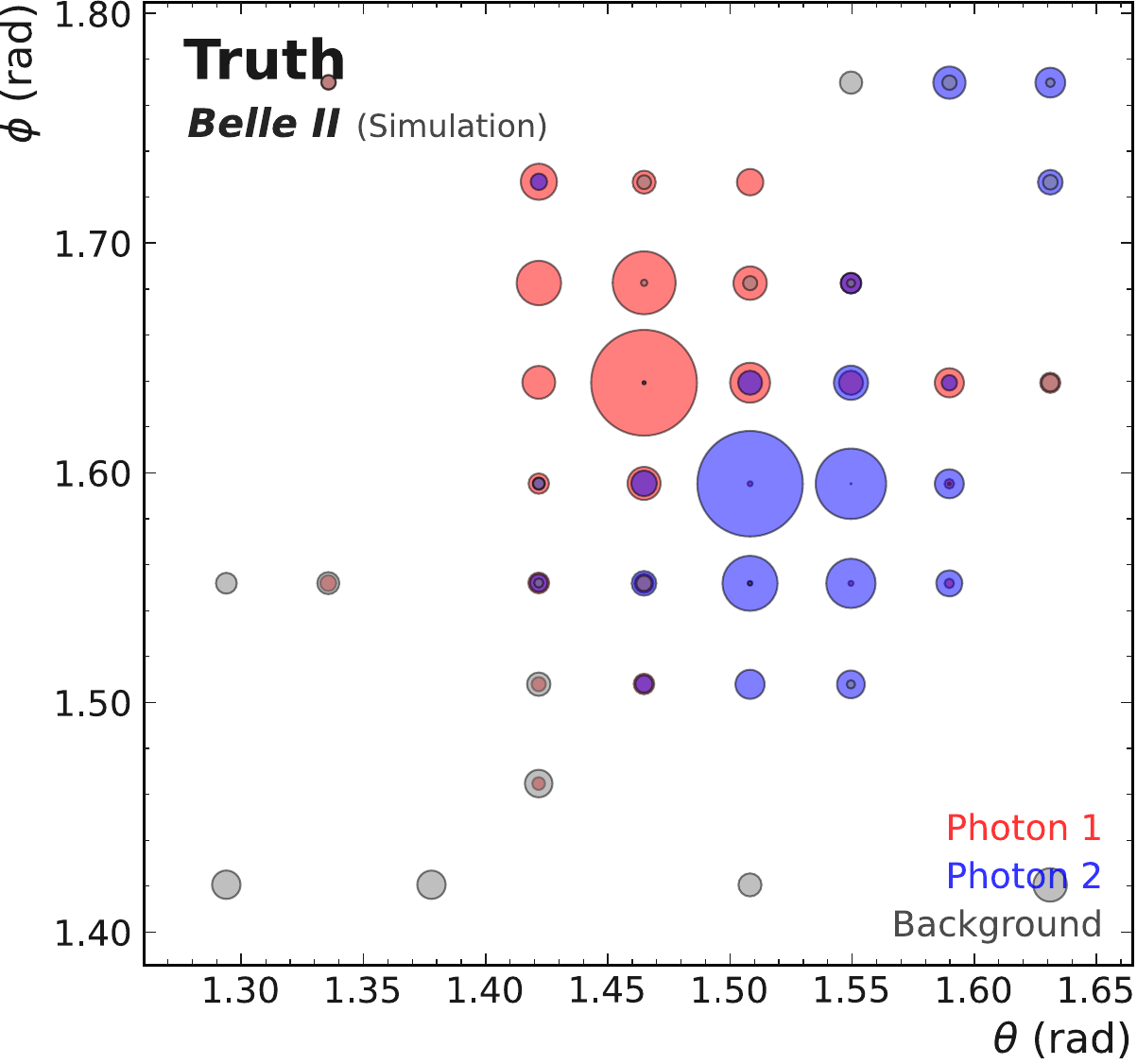}
         \caption{Truth}
         \label{fig:truth_clustering_outlook}
     \end{subfigure}
     \hfill
     \begin{subfigure}[b]{\third\textwidth}
         \centering
         \includegraphics[width=\textwidth]{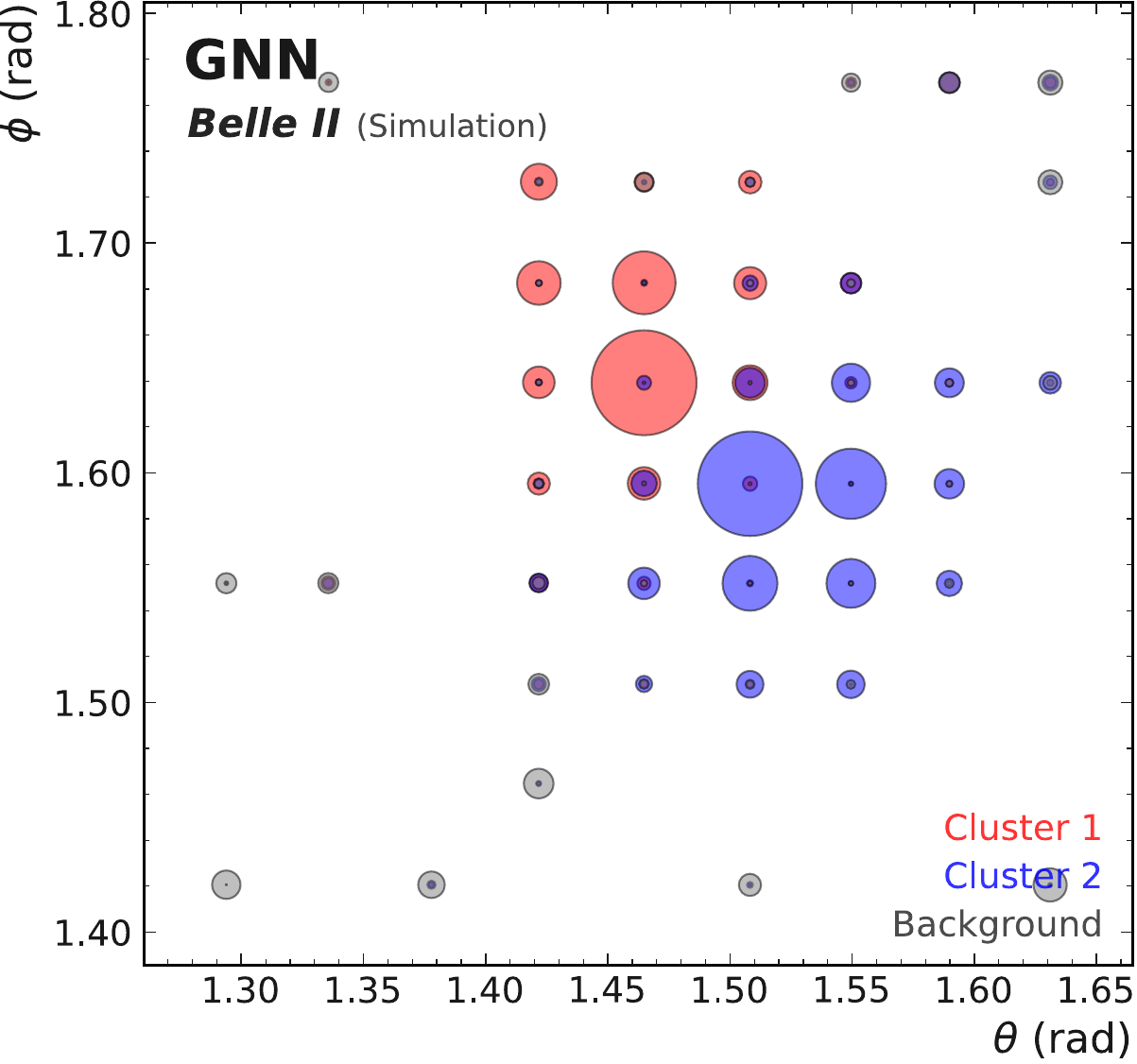}
         \caption{GravNet}
         \label{fig:gravnet_clustering_outlook}
     \end{subfigure}
     \hfill
     \begin{subfigure}[b]{\third\textwidth}
         \centering
         \includegraphics[width=\textwidth]{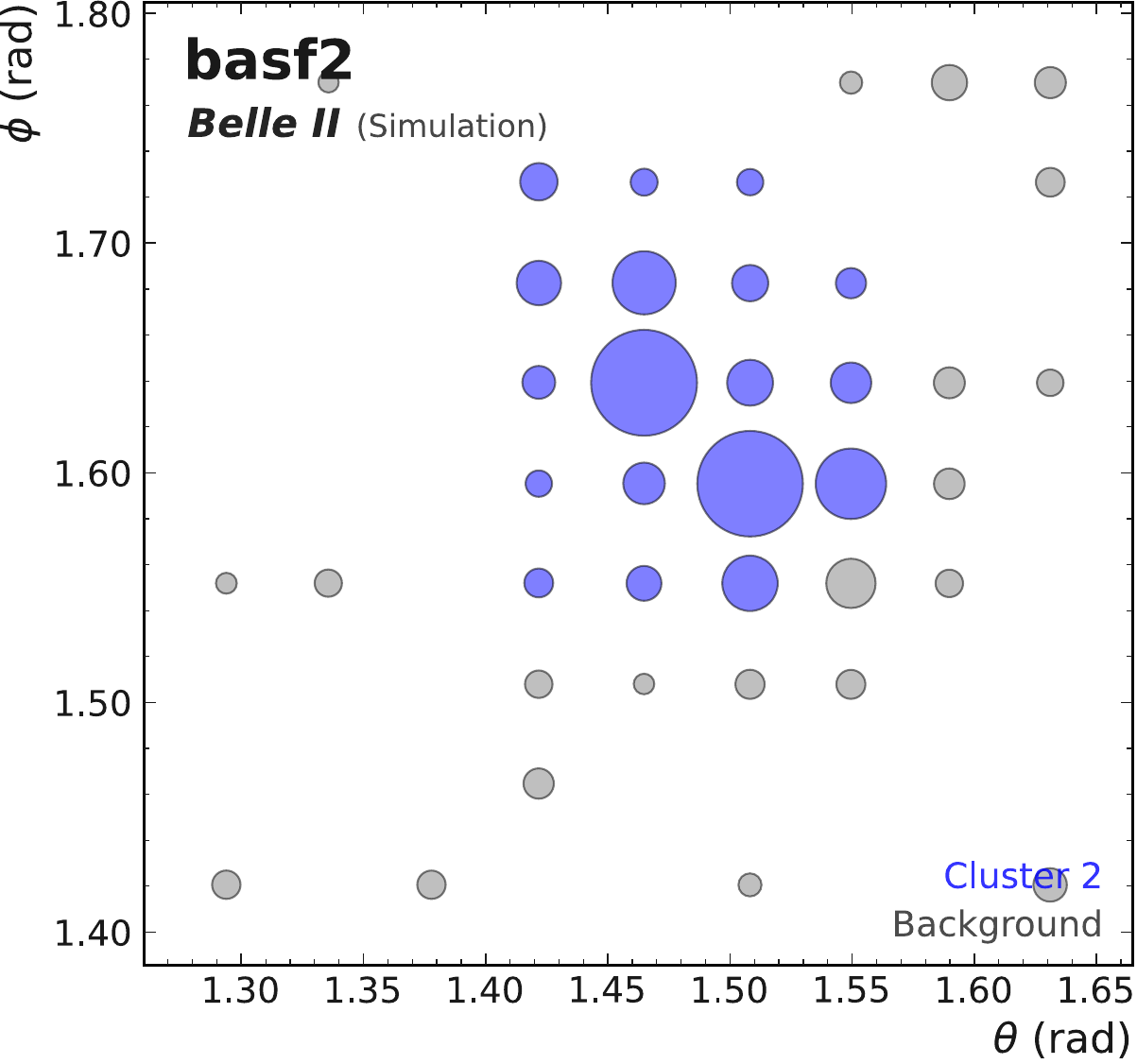}
         \caption{\basf}
         \label{fig:basf2_clustering_outlook}
     \end{subfigure}
        \caption{Comparison of truth energy fractions (\subref{fig:truth_clustering_outlook}), the reconstructed energy fraction by the GNN  (\subref{fig:gravnet_clustering_outlook}), and
the reconstructed energy fraction by \basf (\subref{fig:basf2_clustering_outlook}) for one example event with only one local maximum. 
Colors indicate the fractions belonging to each photon or background. 
The marker centers indicate the crystal centers, the marker area is proportional to the
reconstructed energy in each crystal.}
        \label{fig:clustering_outlook}
\end{figure*}

In this work, we have presented a complete study of a GNN-based fuzzy clustering algorithm for the \belletwo electromagnetic calorimeter.
We have been using a realistic full detector simulation and simulated beam background for low and high luminosity conditions of \belletwo.
The GNN algorithm has been compared to the currently used \basf baseline algorithm.
We find a significantly improved resolution of more than 30\,\% for high beam backgrounds, but also improved performance in reducing the right-side tails of the reconstruction errors that are caused by beam background.
Such significant improvements in photon reconstruction performance directly improve the physics reach of \belletwo for almost all final states with photons, but also analyses that use missing energy information~\cite{Kou:2018nap}.
We also trained different GNNs to separate energy depositions of overlapping photon clusters.
The improvement of the energy resolution is up to 30\,\% for the low energy photon in asymmetric photon pairs.
Any improvement in overlapping photon reconstruction has direct implications for the reconstruction of boosted $\pi^0$ mesons or axion-like particles with couplings to photons~\cite{Belle-II:2020jti}.

While the \basf algorithm strictly reconstructs one cluster for each LM, the GNN algorithm only uses the LMs to center the ROI.
The GNN algorithm can therefore in principle also be used to reconstruct overlapping photons that only produced one LM (Fig.~\ref{fig:clustering_outlook}). 
The extension of the GNN algorithm to such overlapping signatures as well as to charged particles and neutral hadrons will be the focus of follow-up work. 
Future work is also going to address robustness against varying beam backgrounds explicitly, for example by introducing features that are directly sensitive to beam-background levels.

This is the first application of a GNN-based clustering algorithm at \belletwo for a realistic detector geometry and realistic and high beam backgrounds.
This is also the first time that an algorithm has shown to improve the performance of the photon reconstruction by explicitly including timing information on clustering level at \belletwo.

\begin{table*}[!ht]
\caption{$\mathrm{FWHM_{gen}}\times10^{2}$ of one photon with photon energy $E_\gamma^{(1)}$ in dependence of the second photon energy $E_\gamma^{(2)}$ for low beam background for the full detector (barrel and endcaps combined). The uncertainties of the \fwhm for the two algorithms are correlated for each energy interval since they use the same simulated events. The improvement over the \basf baseline algorithm is stated in percent for each energy interval.}
\label{tab:fwhm_low}
\begin{tabularx}{\textwidth}{llXXXX}
\toprule
\makecell[l]{$E_\gamma^{(1)}$ (GeV)\\ $\downarrow$}    & \makecell[l]{$E_\gamma^{(2)}$ (GeV)\\ \hfill$\rightarrow$} & \makecell[rb]{[0.1, 0.2]}& \makecell[rb]{[0.2, 0.5]} & \makecell[rb]{[0.5, 1.0]} & \makecell[rb]{[1.0, 1.5]}  \\
\midrule
\multirow[t]{3}{*}{[0.1, 0.2]}          & GNN    &   \makecell[r]{11.04$\pm$0.79}    &  \makecell[r]{11.98$\pm$0.40}   &   \makecell[r]{11.94$\pm$0.31}     &  \makecell[r]{13.25$\pm$0.34}\\
                                        & \basf      &   \makecell[r]{12.72$\pm$0.80}    &  \makecell[r]{13.93$\pm$0.55}   &   \makecell[r]{14.32$\pm$0.41}     &  \makecell[r]{15.16$\pm$0.48}    \\
                                        & \textbf{Improvement}      &   \makecell[r]{\textbf{15.2\,\%}}     &  \makecell[r]{\textbf{16.3\,\%}}   &   \makecell[r]{\textbf{20.0\,\%}}     &  \makecell[r]{\textbf{14.4\,\%}}  \\
\midrule                         
\multirow[t]{3}{*}{[0.2, 0.5]}          & GNN    &   \makecell[r]{7.38$\pm$0.18}    &  \makecell[r]{7.57$\pm$0.12}   &   \makecell[r]{8.23$\pm$0.09}     &  \makecell[r]{8.38$\pm$0.12}\\
                                        & \basf      &   \makecell[r]{8.48$\pm$0.22}    &  \makecell[r]{8.30$\pm$0.14}   &   \makecell[r]{8.84$\pm$0.12}     &  \makecell[r]{8.96$\pm$0.12}    \\
                                        & \textbf{Improvement}      &   \makecell[r]{\textbf{14.9\,\%}}     &  \makecell[r]{\textbf{9.7\,\%}}   &   \makecell[r]{\textbf{7.5\,\%}}     &  \makecell[r]{\textbf{7.0\,\%}}  \\
\midrule                                        
\multirow[t]{3}{*}{[0.5, 1.0]}          & GNN    &   \makecell[r]{5.22$\pm$0.08}    &  \makecell[r]{5.43$\pm$0.05}   &   \makecell[r]{5.69$\pm$0.04}     &  \makecell[r]{5.89$\pm$0.04}\\
                                        & \basf      &   \makecell[r]{5.58$\pm$0.10}    &  \makecell[r]{5.71$\pm$0.06}   &   \makecell[r]{5.85$\pm$0.05}     &  \makecell[r]{6.17$\pm$0.05}    \\
                                        & \textbf{Improvement}      &   \makecell[r]{\textbf{6.7\,\%}}     &  \makecell[r]{\textbf{5.1\,\%}}   &   \makecell[r]{\textbf{2.8\,\%}}     &  \makecell[r]{\textbf{4.9\,\%}}  \\
\midrule                                        
\multirow[t]{3}{*}{[1.0, 1.5]}          & GNN    &   \makecell[r]{4.24$\pm$0.06}    &  \makecell[r]{4.43$\pm$0.04}   &   \makecell[r]{4.67$\pm$0.03}     &  \makecell[r]{4.77$\pm$0.03}\\
                                        & \basf      &   \makecell[r]{4.55$\pm$0.07}    &  \makecell[r]{4.58$\pm$0.04}   &   \makecell[r]{4.74$\pm$0.04}     &  \makecell[r]{4.85$\pm$0.04}    \\
                                        & \textbf{Improvement}      &   \makecell[r]{\textbf{7.3\,\%}}     &  \makecell[r]{\textbf{3.4\,\%}}   &   \makecell[r]{\textbf{1.4\,\%}}     &  \makecell[r]{\textbf{1.8\,\%}}  \\
\bottomrule
\end{tabularx}
\end{table*}

\begin{table*}[!ht]
\caption{$\mathrm{FWHM_{gen}}\times10^{2}$ of one photon with photon energy $E_\gamma^{(1)}$ in dependence of the second photon energy $E_\gamma^{(2)}$ for high beam background for the full detector (barrel and endcaps combined). The uncertainties of the \fwhm for the two algorithms are correlated for each energy interval since they use the same simulated events. The improvement to the \basf baseline is stated in percent for each energy interval.}
\label{tab:fwhm_high}
\begin{tabularx}{\textwidth}{llXXXX}
\toprule
\makecell[l]{$E_\gamma^{(1)}$ (GeV)\\ $\downarrow$}    & \makecell[l]{$E_\gamma^{(2)}$ (GeV)\\ \hfill$\rightarrow$} & \makecell[rb]{[0.1, 0.2]}& \makecell[rb]{[0.2, 0.5]} & \makecell[rb]{[0.5, 1.0]} & \makecell[rb]{[1.0, 1.5]}  \\
\midrule
\multirow[t]{3}{*}{[0.1, 0.2]}          & GNN     &   \makecell[r]{24.77$\pm$0.83}    &  \makecell[r]{24.10$\pm$0.76}    &   \makecell[r]{24.02$\pm$0.60}       &  \makecell[r]{24.72$\pm$0.63}\\
                                        & \basf       &   \makecell[r]{33.12$\pm$1.08}    &  \makecell[r]{32.82$\pm$1.38}    &   \makecell[r]{31.28$\pm$0.79}       &  \makecell[r]{32.42$\pm$0.88}    \\
                                        & \textbf{Improvement} &   \makecell[r]{\textbf{33.7\,\%}}  &  \makecell[r]{\textbf{36.2\,\%}}   &   \makecell[r]{\textbf{30.3\,\%}}    &  \makecell[r]{\textbf{31.1\,\%}}  \\
\midrule                         
\multirow[t]{3}{*}{[0.2, 0.5]}          & GNN     &   \makecell[r]{13.16$\pm$0.30}    &  \makecell[r]{13.96$\pm$0.20}    &   \makecell[r]{14.17$\pm$0.16}       &  \makecell[r]{14.17$\pm$0.16}\\
                                        & \basf       &   \makecell[r]{17.73$\pm$0.47}    &  \makecell[r]{17.56$\pm$0.31}    &   \makecell[r]{17.62$\pm$0.24}       &  \makecell[r]{16.88$\pm$0.23}    \\
                                        & \textbf{Improvement} &   \makecell[r]{\textbf{34.8\,\%}}  &  \makecell[r]{\textbf{25.8\,\%}}   &   \makecell[r]{\textbf{24.3\,\%}}    &  \makecell[r]{\textbf{19.1\,\%}}  \\
\midrule                                        
\multirow[t]{3}{*}{[0.5, 1.0]}          & GNN     &   \makecell[r]{8.07$\pm$0.12}    &  \makecell[r]{8.56$\pm$0.08}    &   \makecell[r]{8.71$\pm$0.06}     &  \makecell[r]{8.84$\pm$0.06}\\
                                        & \basf       &   \makecell[r]{10.53$\pm$0.19}   &  \makecell[r]{10.77$\pm$0.12}   &   \makecell[r]{10.75$\pm$0.09}    &  \makecell[r]{10.73$\pm$0.08}    \\
                                        & \textbf{Improvement} &   \makecell[r]{\textbf{30.6\,\%}} &  \makecell[r]{\textbf{25.8\,\%}}   &   \makecell[r]{\textbf{23.4\,\%}}   &  \makecell[r]{\textbf{21.4\,\%}}  \\
\midrule                                        
\multirow[t]{3}{*}{[1.0, 1.5]}          & GNN     &   \makecell[r]{6.05$\pm$0.08}    &  \makecell[r]{6.33$\pm$0.05}    &   \makecell[r]{6.42$\pm$0.04}       &  \makecell[r]{6.54$\pm$0.04}\\
                                        & \basf       &   \makecell[r]{7.52$\pm$0.12}    &  \makecell[r]{7.56$\pm$0.07}     &   \makecell[r]{7.60$\pm$0.06}      &  \makecell[r]{7.68$\pm$0.06}    \\
                                        & \textbf{Improvement} &   \makecell[r]{\textbf{24.2\,\%}} &  \makecell[r]{\textbf{19.6\,\%}}   &   \makecell[r]{\textbf{18.3\,\%}}   &  \makecell[r]{\textbf{17.4\,\%}}  \\
\bottomrule
\end{tabularx}
\end{table*}

\backmatter%
\bmhead{Data Availability Statement}%
The datasets generated during and analysed during the current study are property of the \belletwo collaboration and not publicly available.
The instructions and code to replicate the
studies in this paper are available at \cite{wemmer_github, wemmer_zenodo}.

\bmhead{Acknowledgments}
The authors would like to thank the Belle II collaboration for useful discussions and suggestions on how to improve this work.
The authors would like to thank Jan Kieseler for helpful discussions.\\

The training of the models was performed on the TOpAS GPU cluster at the Steinbuch Centre for Computing (SCC) at KIT.
This work is funded by Helmholtz~(HGF) Young Investigators Group VH-NG-1303 and BMBF ErUM-Pro~05H23VKKBA.
I.~Haide is supported by the Landesgraduiertenförderung Baden-Württemberg.\\

\section*{Compliance with ethical standards}

\subsection*{Conflict of interest}
The  authors  declare  that  they  have  no  conflict  of  interest.

%%
%%===========================================================================================%%
\bibliographystyle{unsrt}
\bibliography{sn-bibliography}% common bib file

%% Default %%
%%\input sn-sample-bib.tex%
\end{document}